\documentclass[a4paper,11pt]{article}
\usepackage{amsmath,bm,amssymb,color}

\usepackage{graphicx}
\usepackage{subcaption}
\usepackage{jheppub} 
\usepackage{xspace}

\renewcommand{\tabcolsep}{2em}

\def\Q2{\left(Q^{2}\right)}

\def\d{{\rm d}}

\def\l({\left(}
\def\r){\right)}

\def\d{\hbox{d}}

\def\gapprox{\lower .7ex\hbox{$\;\stackrel{\textstyle >}{\sim}\;$}}
\def\lapprox{\lower .7ex\hbox{$\;\stackrel{\textstyle <}{\sim}\;$}}

\newcommand{\NNLOJET}{NNLO\protect\scalebox{0.8}{JET}\xspace}
\def\ATLASI{ATLAS\,I\xspace}
\def\ATLASII{ATLAS\,II\xspace}
\allowdisplaybreaks

\title{Fiducial cross sections for the four-lepton decay mode in Higgs-plus-jet production up to NNLO QCD}

\author{X.\ Chen$^{a}$, T.\ Gehrmann$^{a}$, E.W.N.\ Glover$^b$, A.\ Huss$^c$}

\affiliation{
$^a$Department of Physics, University of Z\"urich, CH-8057 Z\"urich, Switzerland\\
$^b$Institute for Particle Physics Phenomenology, Department of Physics, University of Durham, Durham, DH1 3LE, UK\\
$^c$Theoretical Physics Department, CERN, 1211 Geneva 23, Switzerland}

\emailAdd{xuan.chen@uzh.ch}
\emailAdd{thomas.gehrmann@uzh.ch}
\emailAdd{e.w.n.glover@durham.ac.uk}
\emailAdd{alexander.huss@cern.ch}

\abstract {The four-lepton decay mode of the Higgs boson allows for a clean kinematic reconstruction, thereby 
enabling precision studies of the Higgs boson properties and of its production dynamics. 
We compute the NNLO QCD corrections to fiducial cross sections relevant to this decay mode in the
gluon-fusion production 
of a Higgs boson in association with a hadronic jet, and study the impact of the QCD corrections on the 
fiducial acceptance factors in inclusive Higgs and Higgs-plus-jet production.
 We investigate in detail 
the different definitions used in the ATLAS and CMS measurements to define the fiducial cross sections. 
Differences in the lepton isolation prescription are found to have a sizeable impact on the higher order corrections to the fiducial acceptance factors.}

\preprint{ ZU-TH 27/19, IPPP/19/45, CERN-TH-2019-083 }

\begin{document}
\maketitle
\flushbottom

\section{Introduction}

The discovery of the Higgs boson~\cite{higgsexp}  initiated a large-scale research program aiming for the precise characterisation
of the Higgs boson properties and of the details of the underlying Higgs mechanism of electroweak symmetry breaking. 
These studies rely largely on measurements of the different Higgs boson production modes and decay channels, which 
are becoming increasingly precise as more and more data is accumulated by the LHC experiments. 
The interpretation of these precision data calls for equally precise theory predictions, ideally taking into account the 
final-state definition used in the experimental measurement, including fiducial cuts on the Higgs boson decay products as well as on 
accompanying objects such as hadronic jets. These predictions are obtained in perturbation theory and become increasingly precise as 
higher orders are included. They play an important role in translating the experimental measurements into so-called simplified template 
cross sections~\cite{YR4}, which are then used in precision studies of Higgs boson properties and Higgs boson couplings. 

The dominant Higgs boson production mode at the LHC is gluon fusion~\cite{ggH}, which is mediated through a closed top quark loop, corresponding to 
a Born process at ${\cal O}(\alpha_s^2)$. This 
process displays very large QCD corrections at next-to-leading order (NLO)~\cite{spira,mtfinite}, which motivated extensive work on 
the derivation of higher order QCD corrections to gluon fusion. These are typically obtained in the limit of infinite top quark mass, which is described by 
an effective field theory coupling the Higgs field to gluons. In this limit, inclusive Higgs boson production in gluon fusion~\cite{ggHnlo,ggHnnlo,hnnlo,ggHn3lo} and its 
rapidity distribution~\cite{hrapn3lo} is described
up to third order in perturbation theory (N3LO), ${\cal O}(\alpha_s^5)$. The leading order  gluon fusion process always produces a Higgs boson at vanishing transverse momentum, 
such that the transverse momentum distribution starts only  at ${\cal O}(\alpha_s^3)$.  This distribution is closely related to Higgs-plus-jet production, which 
has been computed to next-to-next-to-leading order (NNLO, ${\cal O}(\alpha_s^5)$) for infinite top quark mass~\cite{caolaH2,njH,caolaH3,ourHpT},   and to NLO including exact top quark 
dependence~\cite{Jones:2018hbb,Lindert:2018iug,Neumann:2018bsx}. These calculations are either performed for a stable Higgs boson~\cite{caolaH2,njH,Jones:2018hbb,Lindert:2018iug,Neumann:2018bsx}, or focus on the 
$H\to \gamma\gamma$~\cite{caolaH3,ourHpT} or $H\to (2l,2\nu)$~\cite{caolaH3} decay modes. 

The Higgs boson decay mode to four charged leptons (through $H\to ZZ^*$) is particularly well-suited for precision studies~\cite{Aad:2015mxa,Khachatryan:2014kca,Aad:2014tca,Khachatryan:2015yvw,ATLASHto4l,ATLASHto4lconf,CMSHto4l,CMSHto4lconf}, due to the 
clear final state signature and the high momentum resolution that 
can be achieved in its reconstruction. At higher order, predictions for fiducial cross sections in this mode are challenging the quality of the numerical 
integrations, owing to the high dimensionality of the Born-level final state phase space. They have been computed for the inclusive Higgs boson production process to 
NNLO~\cite{hnnlo}, corresponding to NLO-level for Higgs boson production at finite transverse momentum. 

In this paper, we extend the NNLO QCD description (for infinite top quark mass) of gluon-fusion Higgs boson production~\cite{ourHpT} to include Higgs boson decays to 
four charged leptons, thereby providing NNLO-accurate predictions, accounting 
for top quark mass effects through rescaling, for fiducial cross sections in this decay channel. The calculation is described in Section~\ref{sec:setup}. We perform a detailed phenomenological 
study of the recent ATLAS and CMS 13~TeV  measurements~\cite{ATLASHto4l,ATLASHto4lconf,CMSHto4l,CMSHto4lconf} 
of Higgs boson production in the four-lepton decay mode in Section~\ref{sec:result}, investigating in particular the impact of different 
fiducial cross section definitions and lepton isolation prescriptions. Our findings are summarised and discussed in Section~\ref{sec:conc}. 

\section{Setup}
\label{sec:setup}

\subsection{Framework of the calculation}
\label{subsec:Framework}
 Higgs boson production at the LHC mainly proceeds through the gluon-fusion channel, which is mediated by a top quark 
loop.
In the computation of cross sections, this top quark loop can be integrated out, provided that the 
top quark mass is larger than all scales 
involved in the process under consideration. In this limit, one obtains an effective field theory (EFT,~\cite{eft}), whose matching 
 onto full QCD and its renormalisation have been derived to 
 four-loop order~\cite{steinhauser,Kniehl:2006bg}. In the EFT framework, higher-order perturbative corrections can be computed 
 in massless QCD.  

The EFT-based calculation of NNLO QCD corrections to Higgs boson production at finite transverse momentum
has been performed previously for on-shell Higgs boson production~\cite{caolaH2,kirill1,ourH}, as well as for the decay modes to 
two photons~\cite{caolaH3,ourHpT} and two leptons plus two neutrinos~\cite{caolaH3}. It 
involves the following partonic channels:  (a) the two-loop matrix elements of Higgs-plus-three-parton 
processes (double-virtual contributions)~\cite{h3g2l}, (b) the one-loop matrix elements of Higgs-plus-four-parton 
processes (real-virtual contributions)~\cite{h4g1l} and (c) the tree-level matrix elements of Higgs-plus-five-parton processes (double-real contributions)~\cite{h5g0l}.
 All three types of NNLO matrix elements are separately infrared divergent 
 upon integration over the relevant phase spaces 
 and only their sum is finite. Consequently, their implementation requires a
 method to extract and recombine the infrared singularities from each contribution. In our 
 implementation~\cite{ourHpT}, we employ the antenna subtraction method~\cite{ourant}. 
 Our calculation is implemented in a parton-level event generator \NNLOJET.
The calculation of Higgs boson decay to $ZZ^*\rightarrow 4l$ is performed 
in the narrow-width approximation using the 
leading-order matrix elements~\cite{HtoZZto4l}. The implementation of Higgs boson decay in \NNLOJET is 
checked by comparing Born-level predictions for fiducial cross sections for final states with up to three jets with  MCFM~\cite{MCFM7.0} and HNNLO~\cite{hnnlo}.

In our numerical computations, we take the Higgs boson mass of $m_H=125$~GeV and the corresponding decay width of $4.1\times 10^{-3}$~GeV.
The vacuum expectation value is fixed to $v=246.2$~GeV and the top quark mass (appearing in the Wilson coefficient at NNLO) is 173.2~GeV.
We use the PDF4LHC15 parton distribution functions (PDFs)~\cite{nnpdf}
with the value of $\alpha_s(m_Z)=0.118$ and $m_Z=91.1876$~GeV. Note that we systematically use the 
same set of PDFs (PDF4LHC15\_nnlo\_mc) and the same value of $\alpha_s(m_Z)$ for LO, NLO and NNLO predictions. External quarks in initial and final states and final-state leptons are all considered to be massless. 
The factorisation ($\mu_F$) and renormalisation ($\mu_R$) central scales are chosen dynamically on an event-by-event basis as,
\begin{equation}
  \label{eq:scale}
\mu \equiv \mu_R = \mu_F = \frac{1}{2}\sqrt{m_H^2 + (p^{4l}_T)^2},
\end{equation}
where $p^{4l}_T$ is the transverse momentum of the final-state four-lepton system. 
The theoretical uncertainty is estimated by a seven-point scale variation, which  
amounts to varying the renormalisation and factorisation scales
independently by factor of $1/2$ and 2 around $\mu$ to the combinations 
of $(\mu_R,\mu_F)=(\mu/2,\mu/2),(2\mu,2\mu),(\mu,2\mu),(2\mu,\mu),(\mu,\mu/2)$ and $(\mu/2,\mu)$ and taking the envelope of the predictions as error estimate.  

\subsection{Quark mass effects}
\label{subsec:quarkmass}

With the enlargement of the LHC data set, Higgs boson production can be measured over an increasingly large 
kinematic range. 
For example, by combining  measurements of various Higgs boson decay channels, the Higgs boson transverse momentum distribution above 350~GeV could recently be determined  
with an error of about $\pm 25$\%~\cite{Aaboud:2018ezd,CMS:2018hhg}.  At those high transverse
momenta,  the quark loop probes energy scales  comparable to or larger than the top quark mass. Consequently, the EFT description is no longer applicable to the Higgs boson production via gluon fusion.  This implies that top quark mass effects can not be neglected in the large transverse momentum region above $p^H_T\sim m_t$. The exact matrix elements with full top quark mass dependence have been known for some time at Born level (one-loop) for Higgs-plus-jet final states~\cite{Ellis:1987xu,Baur:1989cm}. Recent progress in numerical calculations of two-loop matrix elements~\cite{Li:2015foa,Borowka:2015mxa,Borowka:2017idc} enabled the 
calculation of the NLO QCD corrections to Higgs-plus-jet final states with full top quark mass 
dependence~\cite{Jones:2018hbb}. Analytical calculations of approximated massive two-loop matrix elements in~\cite{Melnikov:2016qoc,Melnikov:2017pgf,Kudashkin:2017skd} allowed 
detailed NLO phenomenology studies of top mass effects in the large transverse momentum limit~\cite{Neumann:2018bsx} and of top-bottom interference effects in Higgs-plus-jet production at the LHC~\cite{Lindert:2017pky,Caola:2018zye}.

For fully consistent predictions of Higgs boson production at large transverse momentum, 
one would ideally derive 
the NNLO QCD corrections involving matrix elements with full top quark mass dependence, which would 
imply computing three-loop four-point amplitudes with an internal mass. 
Due to the complexity of computing those amplitudes, such calculations are not feasible at present. Instead of the full calculation, one can apply re-weighting procedures to approximate quark mass effects in the higher-order EFT predictions. 
For inclusive observables such as the Higgs boson production cross section, the NLO top quark mass corrections are observed~\cite{HNLOmass} to be well-approximated by re-weighting the NLO EFT cross section by the ratio between the full and EFT predictions at LO:
\begin{equation}
 R_{\text{LO}}=\sigma^\text{M}_{\text{LO}}/\sigma^{\text{EFT}}_{\text{LO}}  ,
\end{equation}
where $\sigma^\text{M}_{\text{LO}}$ includes the exact mass dependence of top quark loops. For differential observables such as the Higgs boson transverse momentum distribution, comparing predictions at LO and NLO between the EFT and including the full top mass dependence, the work in~\cite{Jones:2018hbb} has demonstrated a rather flat ratio for NLO over LO predictions at large transverse momentum. This indicates that the NLO top mass effects can be estimated by a multiplicative rescaling of the NLO EFT predictions:
\begin{equation}
  \label{eq:NLOpTrewt}
 \frac{\d \sigma^\text{M}_{\text{NLO}}}{\d p^H_T}\approx R_{\text{LO}}(p^H_T)\bigg( \frac{\d \sigma^{\text{EFT}}_{\text{NLO}}}{\d p^H_T}\bigg) ,
\end{equation}
where $R_{\text{LO}}(\mathcal O)$ is the generalised LO re-weighting factor for the differential distribution of an observable $\mathcal O$
\begin{equation}
  \label{eq:Rdiff}
  R_{\text{LO}}(\mathcal O)= \bigg(\frac{\d \sigma^{\text{M}}_{\text{LO}}}{\d \mathcal O}\bigg)\bigg/\bigg(\frac{\d \sigma^{\text{EFT}}_{\text{LO}}}{\d \mathcal O}\bigg),
\end{equation}
which is computed with the same fiducial cuts as the cross section under consideration.
The multiplicative re-weighting procedure which is defined by \eqref{eq:NLOpTrewt} has been referred to as the EFT$\otimes$LOM approximation in~\cite{ourHpT}, where the differential cross sections in Higgs-plus-jet production in 
the   di-photon decay channel were studied up to NNLO EFT$\otimes$LOM accuracy. To estimate the quark mass effects in the Higgs boson to four-lepton decay channel, we systematically apply this multiplicative re-weighting procedure to NLO and NNLO EFT predictions:
\begin{equation}
  \label{eq:Hto4lrewt}
 \frac{\d \sigma_{\text{N(N)LO}\otimes \text{LOM}}}{\d \mathcal O}= R_{\text{LO}}(\mathcal O)\bigg( \frac{\d \sigma^{\text{EFT}}_{\text{N(N)LO}}}{\d \mathcal O}\bigg).
\end{equation}
With the recent results~\cite{Jones:2018hbb} for the exact top quark mass dependence of the NLO corrections
 in Higgs-plus-jet production at the LHC, this procedure could in the future be refined through an 
 NLO re-weighting factor $R_{\text{NLO}}(\mathcal O)$ which would define an EFT$\otimes$NLOM accuracy,
 similar to the prescription used for double Higgs boson production~\cite{Grazzini:2018bsd}. 

The mass effects from bottom and charm quarks are numerically smaller than those from top quarks. Nevertheless, one can include those effects by adding the corresponding quark loops in the numerator of Eq.~\eqref{eq:Rdiff}. In this paper, all re-weighting factors $R_{\text{LO}}$ or $R_{\text{LO}}(\mathcal O)$ include massive charm, bottom and top loops in $\sigma^\text{M}_{\text{LO}}$ or $\d \sigma^\text{M}_{\text{LO}}$ with masses of {$m_t = 173.2$~GeV, $m_b =4.18$~GeV, $m_c =1.275$~GeV} while keeping other light quarks massless. 

\begin{figure}
    \centering
    \begin{subfigure}[b]{0.4\textwidth}
        \includegraphics[width=\textwidth]{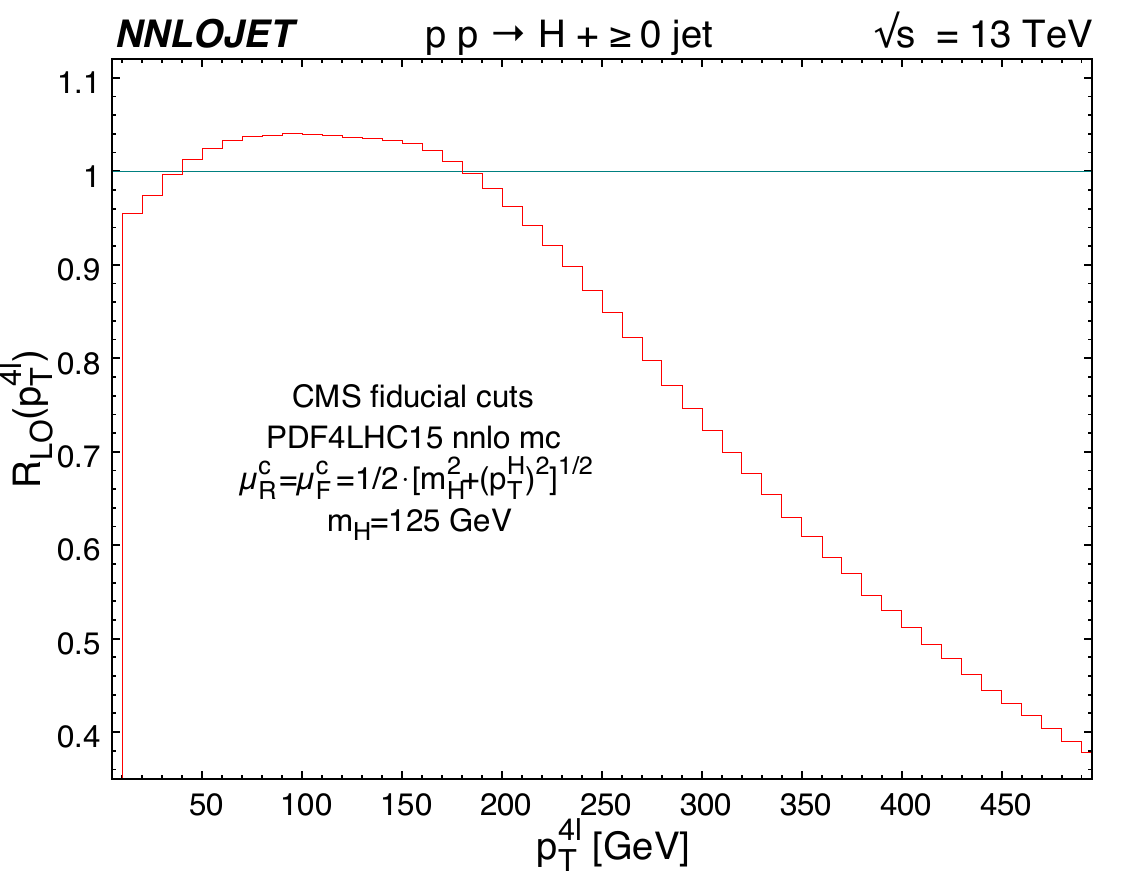}
    \end{subfigure}
    \begin{subfigure}[b]{0.4\textwidth}
        \includegraphics[width=\textwidth]{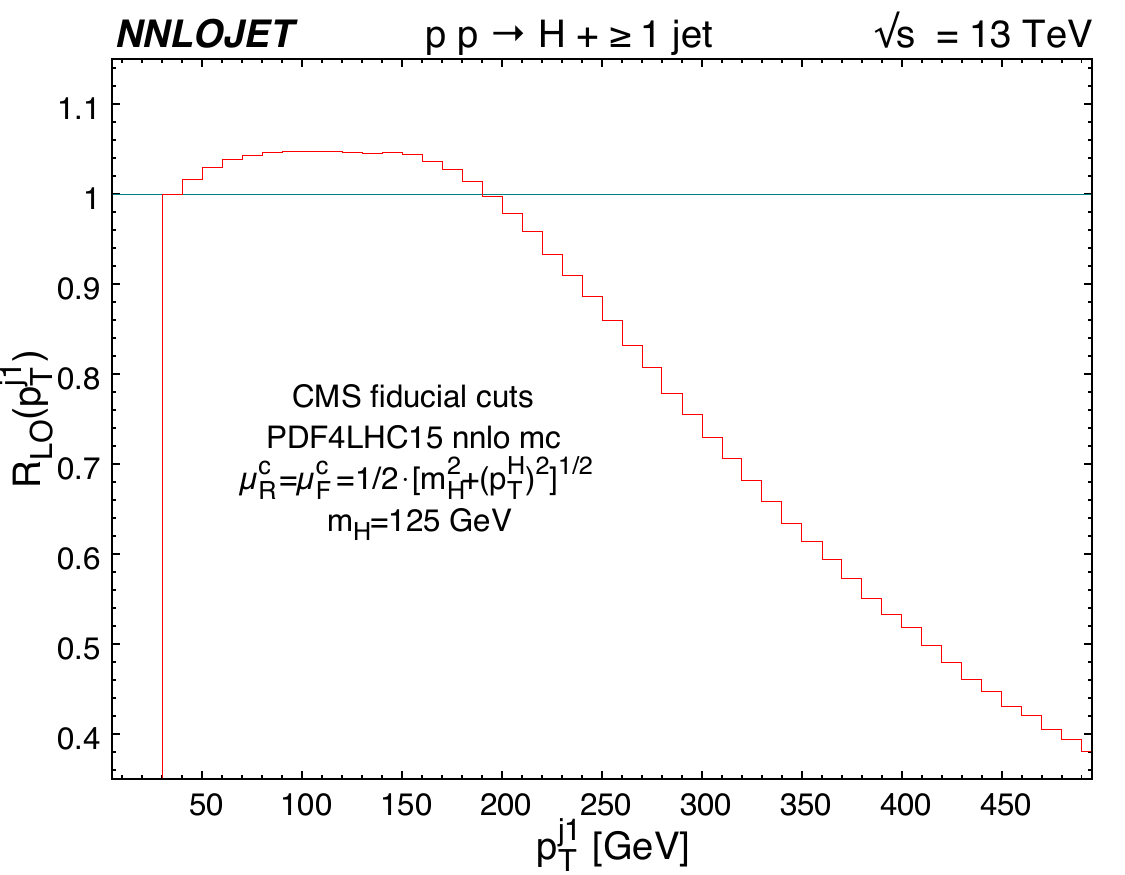}
    \end{subfigure}
    \caption{The scaling factor $R_{\text{LO}}(p^{4l}_T)$ (left) and $R_{\text{LO}}(p^{j1}_T)$ (right) at leading order with exact quark mass dependence.}
    \label{fig:Rfactor}
\end{figure}

To illustrate the corrections to EFT predictions with massive top, bottom and charm quarks we apply fiducial cuts based on the CMS measurements~\cite{CMSHto4l,CMSHto4lconf} with $H\rightarrow ZZ^*\rightarrow 4l$ (+ jets) final states  (see Section~\ref{sec:fiducial} below for details) as an example. Figure~\ref{fig:Rfactor} shows $R_{\text{LO}}(p^{4l}_T)$ for Higgs boson production as a function of $p^{4l}_T$ and $R_{\text{LO}}(p^{j1}_T)$ for Higgs-plus-jet production as a function of $p^{j1}_T$ (transverse momentum of the leading jet). We observe that the exact quark mass dependence leads to a mild enhancement (up to about 4.1\% for $R_{\text{LO}}(p^{4l}_T)$ and up to about 4.8\% for $R_{\text{LO}}(p^{j1}_T)$) in the transverse momentum region below $m_t$. For $(p_T^{4l}, p_T^{j1})> m_t$, $R_{\text{LO}}(p^{4l}_T)$ and $R_{\text{LO}}(p^{j1}_T)$ falls off steeply with increasing transverse momentum as the top quark mass in the loop starts to be resolved.

\subsection{Definition of the fiducial cross sections}
\label{sec:fiducial}
In order to facilitate a direct comparison with the differential cross sections measured by CMS (2016 data~\cite{CMSHto4l} and 2016-18 data~\cite{CMSHto4lconf}) and ATLAS (2016 data~\cite{ATLASHto4l} and
 2016-17 data~\cite{ATLASHto4lconf}), we apply fiducial event selection cuts in the 
Higgs-to-four-lepton decay mode (with or without associated jets), which are summarized in Table~\ref{tab:allthree}. The fiducial selection criteria are slightly different between the two ATLAS measurements in~\cite{ATLASHto4l} and~\cite{ATLASHto4lconf}, while the two CMS measurements adopt the same fiducial cuts in both their analysis of a partial set~\cite{CMSHto4l} and the full set~\cite{CMSHto4lconf} of the Run II data.  For the $H\rightarrow ZZ^*\rightarrow 4l$ decay mode, the measured lepton pairs come in three categories: $2e2\mu$, $4e$ or $4\mu$ while $\tau$ lepton pairs are excluded. Same-flavour-opposite-sign (SFOS) lepton pairs are identified to constrain the invariant mass of $Z$ boson candidates. In the $2e2\mu$ case, there are only two SFOS pairs, while in the $4e$ or $4\mu$ case, there are four possible assignments of SFOS pairs. The SFOS lepton pair with invariant mass closest to $m_Z$ is labelled as the $Z_1$ candidate, while the remaining SFOS lepton pair is labelled as the $Z_2$ candidate. The $H\rightarrow ZZ^*$ decay process creates two off-shell $Z$ bosons, and the CMS~\cite{CMSHto4l,CMSHto4lconf} and ATLAS~\cite{ATLASHto4l,ATLASHto4lconf} measurements use different
invariant mass cuts on $m_{Z_1}$ and $m_{Z_2}$, as well as on the 
four-lepton invariant mass $m_{4l}$. 
Our calculation includes all these experimental cuts, with the exception of the four-lepton 
invariant mass, which is in our case always fixed to be the on-shell value of $m_H$. Owing to the small width of the 
Higgs boson, this approximation is justified for the ATLAS and CMS cuts on $m_{4l}$. Off-shell Higgs boson 
production becomes sizeable in this decay mode only for invariant masses above $2m_Z$, which are vetoed by 
the experimental cuts. 
\begin{table}
\begin{center}
\begin{tabular}{|l|c|c|c|}
  \hline
  & CMS & \ATLASI & \ATLASII\\
  \hline
\multicolumn{4}{|l|}{Lepton Kinematics}\\
\hline
1st lepton $p_T^{l_1}$ (GeV) & $>20$  & $>20$  &$>20$\\
2nd lepton $p_T^{l_2}$ (GeV)& $>10$  & $>15$   &$>15$\\
3rd lepton $p_T^{l_3}$ (GeV)& --- & $>10$  & $>10$\\
lepton $p_T^{e(\mu)}$ (GeV)& $>7 (5)$ & $>7(5)$ &$>5$\\
Rapidity $|y^{e(\mu)}|$ & $<2.5(2.4)$ & $<2.47(2.7)$ &$<2.7$\\
\hline
\multicolumn{4}{|l|}{Lepton Isolation}\\
\hline
Cone size $ R^l$ & $0.3$ & --- & ---\\
$\sum_i p^i_T / p^l_T$ ( $i \in  R^l$)  & $ < 0.35$ & --- & ---\\
 $\Delta R^{\text{SF(DF)}}(l_i,l_j)$ & $ >0.02$ & $>0.1 (0.2)$ & $>0.1$\\
\hline
\multicolumn{4}{|l|}{Invariant Mass (GeV)}\\
\hline
 $Z_1$ candidate $m_{Z_1}$  & $[40, 120]$ & $[50, 106]$ &$[50, 106]$ \\
 $Z_2$ candidate $m_{Z_2}$  & $[12, 120]$ & $[12, 115]$ &$[12, 115]$ \\
 $m_{l^+ l^{'-}}$ (SF+DF) & $> 4$ & --- & ---\\
 $m_{l^+ l^{-}}$ (SF)& --- & $> 5$ & $> 5$\\
Four leptons $m_{4l}$ & 125 & 125 & 125\\
\hline
\multicolumn{4}{|l|}{Jet Definition}\\
\hline
Algorithm & anti-$k_T$ & anti-$k_T$ & anti-$k_T$ \\
Cone size R & 0.4 & 0.4 & 0.4\\
$p_T^{j}$ (GeV) & $>30$ & $>30$ &  $>30$\\
Rapidity $|y^{j}|$ & $<2.5$ & $<4.4$ & $<4.4$ \\
 $\Delta R(j,e(\mu))$ & --- &$>0.2 (0.1)$ & $>0.1$\\
\hline
\end{tabular}
\end{center}
\caption{Fiducial cuts for final state leptons and jets in the calculations of \NNLOJET (based on the measurements of CMS~\protect{\cite{CMSHto4l,CMSHto4lconf}}, \ATLASI~\protect{\cite{ATLASHto4l}} and \ATLASII~\protect{\cite{ATLASHto4lconf}}). 
Jet definitions do not apply to the inclusive transverse momentum distributions. The prescriptions for lepton-jet separation differ considerably between the 
two experiments, as described in detail in the text. For ATLAS, they are part of the jet definition, while for CMS they are part of the 
lepton isolation. 
\label{tab:allthree}}
\end{table}

The jet-related fiducial cuts and jet algorithm are only applied for the 
Higgs+jet cross sections. CMS~\cite{CMSHto4l,CMSHto4lconf} applies the standard anti-$k_T$ jet algorithm 
for jet clustering and jet counting. ATLAS~\cite{ATLASHto4l,ATLASHto4lconf} uses the anti-$k_T$ jet algorithm for jet clustering while applying an additional criterion for the jet counting: 
In order to avoid ambiguities with the potential assignment of leptons to jets, the ATLAS measurements~\cite{ATLASHto4l,ATLASHto4lconf} remove any jet candidate within $\Delta R(j,l)$ for each final-state
lepton. Note that this criterion only affects the jet counting but does not remove the event from any of 
the histograms such as the Higgs boson transverse momentum distribution. Besides this jet removal, the 
ATLAS measurements do not apply lepton isolation cuts. In contrast, the CMS measurement~\cite{CMSHto4l,CMSHto4lconf}
requires the scalar sum of the transverse momenta of all partons within a cone size of $R^l$ around each lepton to be smaller than 35\% of the lepton's transverse momentum ($\sum_i p^i_T / p^l_T\ <$   0.35 for $i \in  R^l$). If this criterion is 
not fulfilled, the lepton is not considered as a lepton in the list of final state particles. In the theoretical 
 calculation, this implies the removal of the event, since less than four leptons are identified. 
The final-state leptons are also required to be separated from each other by $\Delta R(l_i,l_j)$ for any $i\ne j$ in all four measurements~\cite{CMSHto4l,CMSHto4lconf,ATLASHto4l,ATLASHto4lconf}. The 
ATLAS measurement~\cite{ATLASHto4l} further uses a different cone size for 
$\Delta R(l_i,l_j)$ in the case of same-flavour (SF) and different-flavour (DF) lepton pairs. 
The full set of the fiducial cuts is summarised 
in Table~\ref{tab:allthree}. We will see in Section~\ref{sec:result} below, how the different cuts and isolation requirements 
affect the perturbative stability of the theoretical predictions.

\subsection{Fiducial and inclusive cross sections}
\label{sec:fid-to-inc}

The experimental measurements of fiducial cross sections are used frequently to reconstruct 
simplified template  cross sections, which are kinematic distributions that are corrected to stable Higgs boson production. 
These simplified template cross sections play an important role in 
discriminating different Higgs boson production processes, and allow detailed studies of Higgs boson couplings. 
 In this reconstruction, the measured fiducial cross sections are 
multiplied with acceptance factors (determined from theory or simulation) to extrapolate them to a fully inclusive 
acceptance of the Higgs boson decay products (which we abbreviate as inclusive cross sections in the following). These acceptance  factors are often based on leading-order theory
predictions, since precision calculations for fiducial cross sections for complicated high-multiplicity 
decay channels are only gradually becoming available. 

Our calculations for fiducial cross sections in Higgs boson and Higgs-plus-jet production in the 
$H\rightarrow ZZ^*\rightarrow 4l$ decay mode provide an excellent testing ground for the perturbative stability of these 
acceptance factors. We distinguish between  ${\mathcal X}=H\rightarrow ZZ^*\rightarrow 4l$, denoting 
the final state with fiducial cuts and ${\mathcal X}=H$ denoting Higgs boson production inclusive over its decay products.

By normalising to the respective leading-order predictions, one obtains the commonly used $K$-factor, where all 
multiplicative mass corrections cancel:
\begin{equation}
  \label{eq:NNLOtoLOK}
  K_{\text{N(N)LO}} (\mathcal O)=\frac{\d \sigma^{\mathcal X (+ jet)}_{\text{N(N)LO$\otimes$ LOM}}/\d \mathcal O}{\d \sigma^{\mathcal X (+ jet)}_{\text{LOM}}/\d \mathcal O}= \frac{\d \sigma^{\mathcal X (+ jet)}_{\text{N(N)LO}}/\d \mathcal O}{\d \sigma^{\mathcal X (+ jet)}_{\text{LO}}/\d \mathcal O}.
\end{equation}
We define semi-inclusive differential distributions (no fiducial cuts on Higgs boson decay observables) by using differential cross sections of inclusive Higgs boson production multiplied by the leading-order Higgs-to-four-lepton branching ratio: 
 \begin{equation}
  \label{eq:seminc}
  \frac{\d \tilde{\sigma}^{H (+jet)}_{\text{FO}}}{\d \mathcal O}= \frac{\d \sigma^{H (+jet)}_{\text{FO}}}{
    \d \mathcal O}\times(\text{BR}_{2e2\mu} +\text{BR}_{4\mu} + \text{BR}_{4e}  )\;,
\end{equation}
 with FO being LOM, NLO$\otimes$LOM or NNLO$\otimes$LOM. For $m_H=125$~GeV the branching 
 ratios of the Higgs boson decay are $\text{BR}_{2e2\mu}=5.8\times 10^{-5}$ for
   two different flavour lepton pairs  and $\text{BR}_{4e}=\text{BR}_{4\mu}=3.2\times 10 ^{-5}$ for 
 two identical flavour lepton pairs. The total Higgs boson decay width is $4.1\times 10 ^{-3} $~GeV.
 
The ratio of fiducial and semi-inclusive cross section defines the acceptance correction factor:
\begin{equation}
  \label{eq:accept}
  A_{\text{FO}} (\mathcal O)=\frac{\d \sigma^{H\rightarrow ZZ^*\rightarrow 4l (+ jet)}_{\text{FO}}/\d \mathcal O}{\d \tilde{\sigma}^{H (+ jet)}_{\text{FO}}/\d \mathcal O}.
\end{equation}
This factor can be computed perturbatively to FO = (LO, NLO, NNLO), thereby testing its stability under higher-order corrections and its 
sensitivity to the cuts that define the fiducial cross section.

\section{Results}
\label{sec:result}

Higgs boson production with the subsequent decay to four leptons via two off-shell Z bosons has been measured by ATLAS~\cite{ATLASHto4l,ATLASHto4lconf} and CMS~\cite{CMSHto4l,CMSHto4lconf}, based on the data taken at 13~TeV.  
Both experiments perform measurements in fiducial phase space regions, on one hand, to ensure that all final state observables (jets, leptons) are within the detector coverage, on the other hand, to apply lepton isolation algorithms that distinguish between leptons from Z boson decay and those from electroweak fragmentation of hadrons within jets.
The measured fiducial total and differential cross sections for the four-lepton final states (plus jets) receive contributions from all possible Higgs boson production channels at the LHC. 
In this section we derive NNLO-accurate parton-level predictions for the Higgs boson production via gluon-gluon fusion including the decay to four leptons in association with a jet and investigate 
the effect of the lepton fiducial cuts and lepton isolation requirement on the acceptance factors and their perturbative stability. Sub-dominant Higgs production channels (vector boson fusion, associated production with a vector boson and top anti-top quark fusion) become important at the 10\% level. They also become dominant at large Higgs transverse momentum region beyond 600~GeV. In the comparison to current measurements from ATLAS and CMS, we do not include them in the theory predictions.

\subsection{Acceptance factors for Higgs simplified template cross sections at the LHC}
\label{sec:accept}
For Higgs-boson and Higgs-boson-plus-jet production, simplified template cross sections that are differential in the Higgs kinematics are
 reconstructed from fiducial cross sections in certain Higgs boson decay channels. 
 This is achieved through acceptance factors, defined in Section~\ref{sec:fid-to-inc}, to extrapolate fiducial cross sections defined on the Higgs boson decay products to  inclusive phase space. 

\begin{figure}
  \centering
    \begin{subfigure}[b]{0.4\textwidth}
        \includegraphics[width=\textwidth]{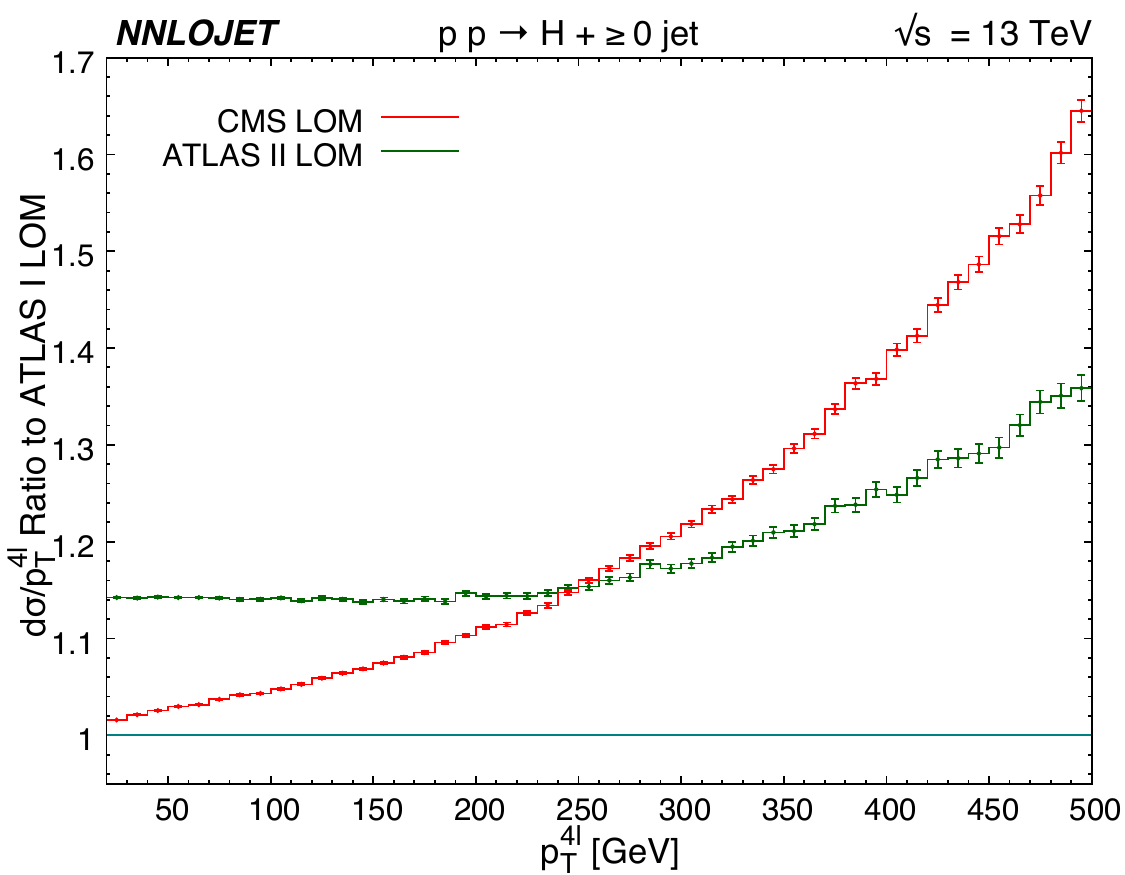}
    \end{subfigure}
\caption{Leading order (with full quark mass effects) four-lepton $p_T$ distributions  at central scale using different fiducial cuts and normalised with respect to the \ATLASI~\cite{ATLASHto4l} fiducial cuts listed in Table~\ref{tab:allthree}. }\label{fig:LOshape}
\end{figure}

The shape of fiducial distributions in Higgs-related kinematic variables may be influenced by the fiducial selection cuts applied to the decay products. This is illustrated 
in Fig.~\ref{fig:LOshape}, which compares  the Higgs boson transverse momentum distribution in the four-lepton final state for the different analyses defined in Table~\ref{tab:allthree}, using 
\ATLASI~\cite{ATLASHto4l} as normalization. At low transverse momentum CMS and \ATLASI coincide, while the \ATLASII fiducial cross section is slightly larger due to 
the larger rapidity acceptance for the leptons. At higher transverse momenta, both CMS and \ATLASII fiducial cuts retain more events than the \ATLASI cuts. The stronger increase for  
the CMS fiducial cuts can be traced back the less strict requirements on the transverse momenta of the sub-leading leptons, which are produced in the decay of highly boosted $Z$-bosons. 

\begin{figure}
    \centering
    \begin{subfigure}[b]{0.32\textwidth}
        \includegraphics[width=\textwidth]{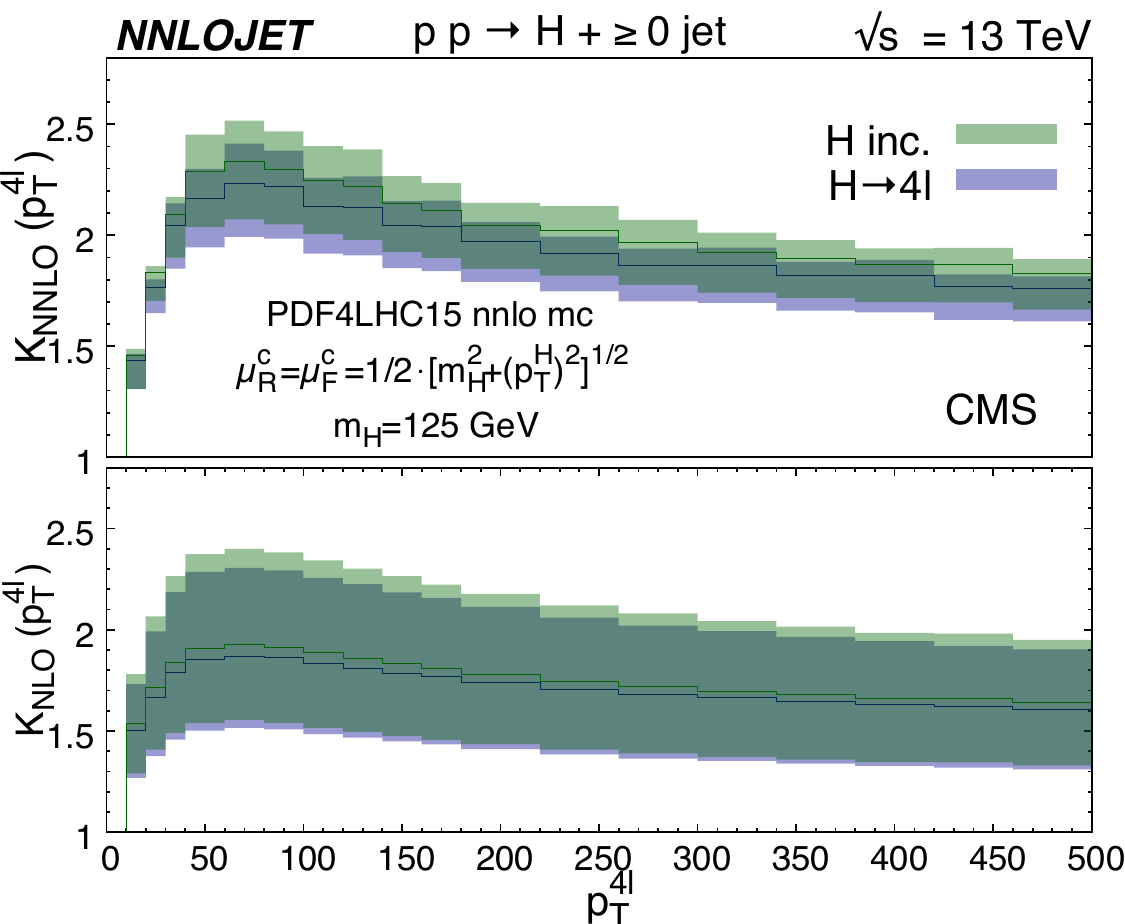}
        \label{fig:CMS-ptnorm}
    \end{subfigure}
    \begin{subfigure}[b]{0.32\textwidth}
        \includegraphics[width=\textwidth]{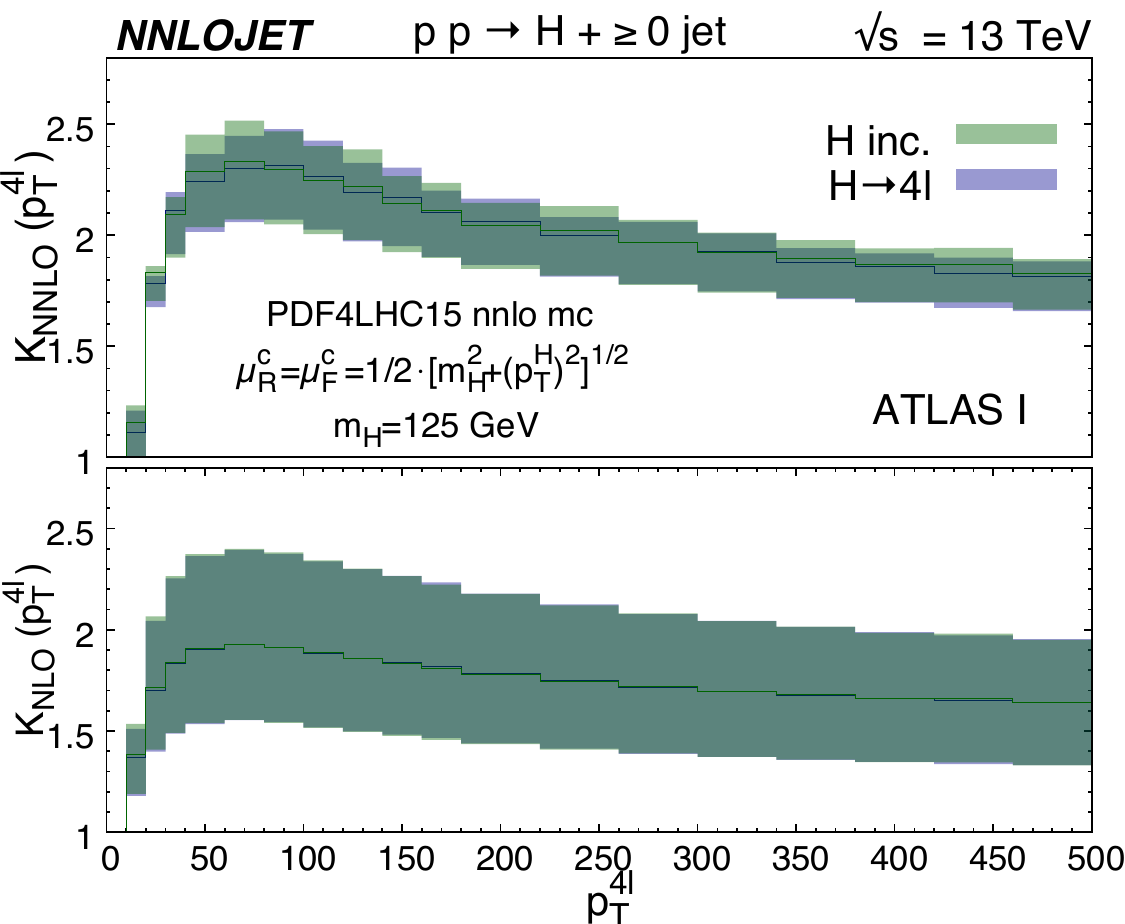}
        \label{fig:ATLAS1-ptnorm}
    \end{subfigure}
    \begin{subfigure}[b]{0.32\textwidth}
        \includegraphics[width=\textwidth]{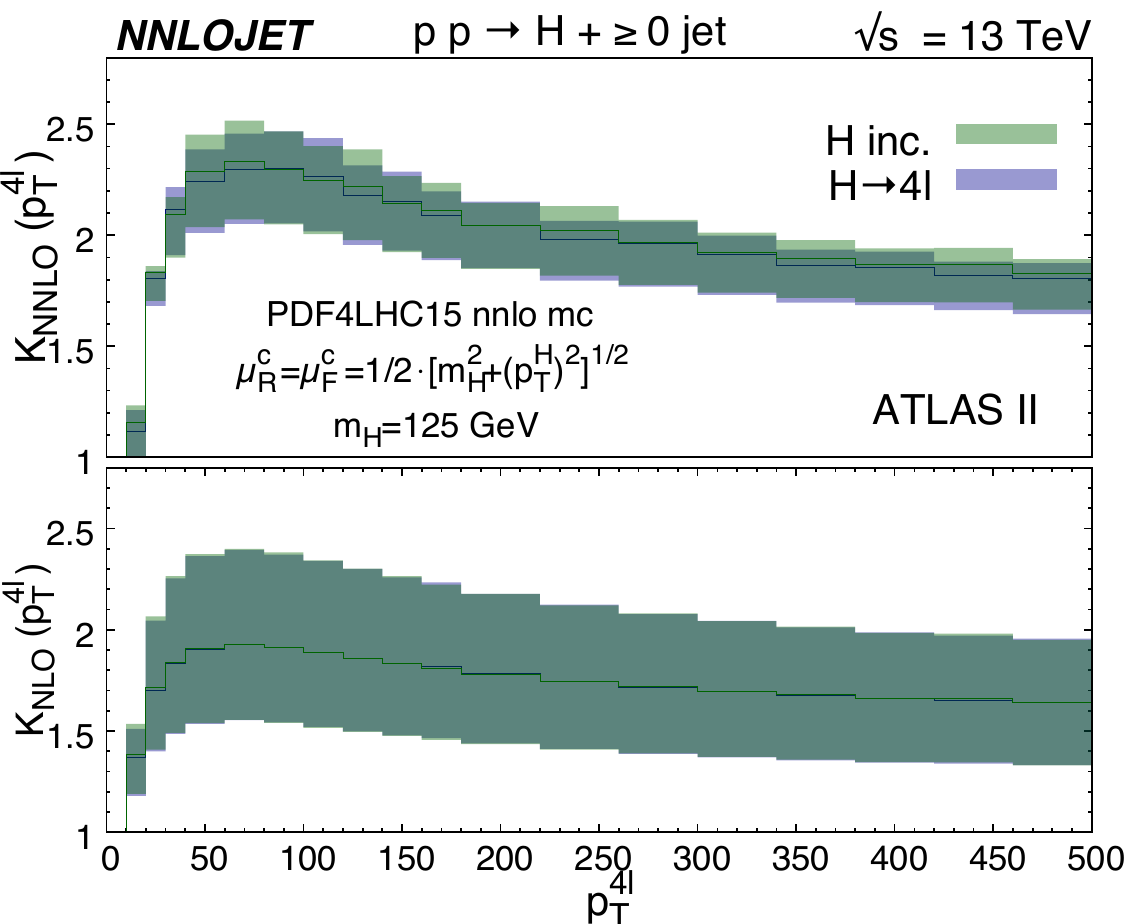}
        \label{fig:ATLAS2-ptnorm}
    \end{subfigure}
    \caption{Transverse momentum distributions of the inclusive Higgs boson or four-lepton system from Higgs boson decay for CMS (left), \ATLASI (centre) and \ATLASII (right) cuts normalised by the corresponding LO distributions.}\label{fig:KfactorHpT}
\end{figure}

Figure~\ref{fig:KfactorHpT} displays the $K$-factors defined in Eq.~\eqref{eq:NNLOtoLOK} for the Higgs boson transverse momentum distributions. The blue curves correspond to 
the $K$-factors for the fiducial four-lepton cross sections (defined in Table~\ref{tab:allthree}), while the green ones indicate the $K$-factors for inclusive Higgs boson production (which are identical in all 
three frames).  
Scale variation bands are obtained by fixing the denominator of the $K$-factors at the central scale while varying the numerator using the seven-point scale variation introduced in Section~\ref{subsec:Framework} above.  
We observe large $K$-factors for the inclusive distribution as well as for all three sets of fiducial cuts at NLO and NNLO. 
The scale variations are reduced going from NLO to NNLO by more than a factor of two. 
For the \ATLASI and \ATLASII analyses, the fiducial $K$-factors at NLO and NNLO are almost identical to the inclusive ones.  
In contrast, deviations between inclusive and fiducial $K$-factors are clearly visible for the CMS fiducial cuts. Already at NLO, the inclusive $K$-factor is about 5\% larger than the fiducial one, and 
this discrepancy increases to about 9\% at NNLO. The lepton isolation prescription applied by CMS~\cite{CMSHto4l,CMSHto4lconf} is the main source of this deviation.   We recall that 
the CMS analysis rejects leptons that are accompanied by hadrons if the sum of hadronic transverse momenta in a cone of size 0.3 around the lepton direction exceeds 35\% of the lepton transverse 
momentum. The increasing amount of parton radiation at higher orders covers more and more final state phase space. This 
leads to the rejection of a larger number of leptons (which passed all other lepton fiducial cuts) at each order,  thereby explaining the 
reduction of the $K$-factor compared to the inclusive case. ATLAS applies the lepton isolation in a completely different manner,  by rejecting jets that 
are found within a cone of radius of  0.1 or 0.2 around the final state leptons. Unlike in the CMS prescription, the lepton is retained, and the event is kept in the analysis. As a consequence, the 
ATLAS fiducial $K$-factors are much closer to the inclusive ones at either perturbative order. 

\begin{figure}
    \centering
    \begin{subfigure}[b]{\textwidth}
\centering
        \includegraphics[width=0.32\textwidth]{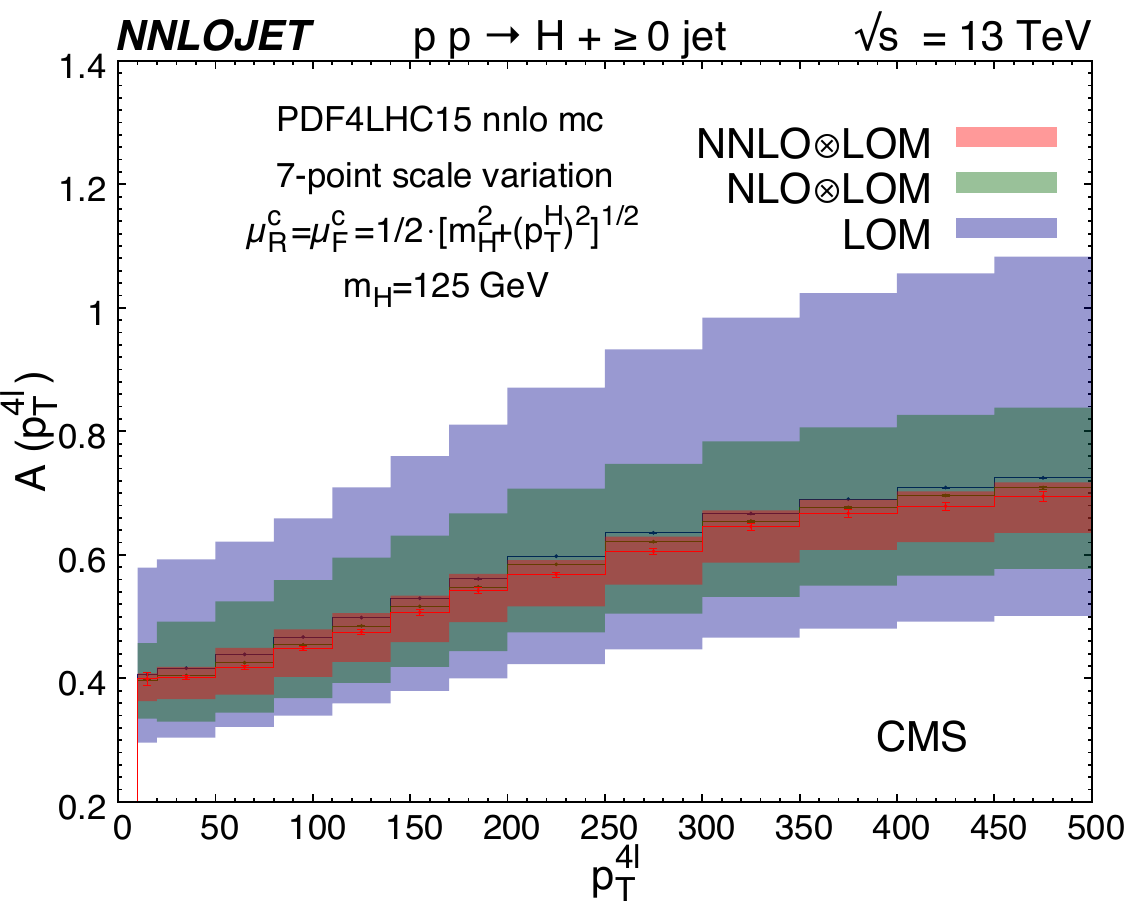}
        \includegraphics[width=0.32\textwidth]{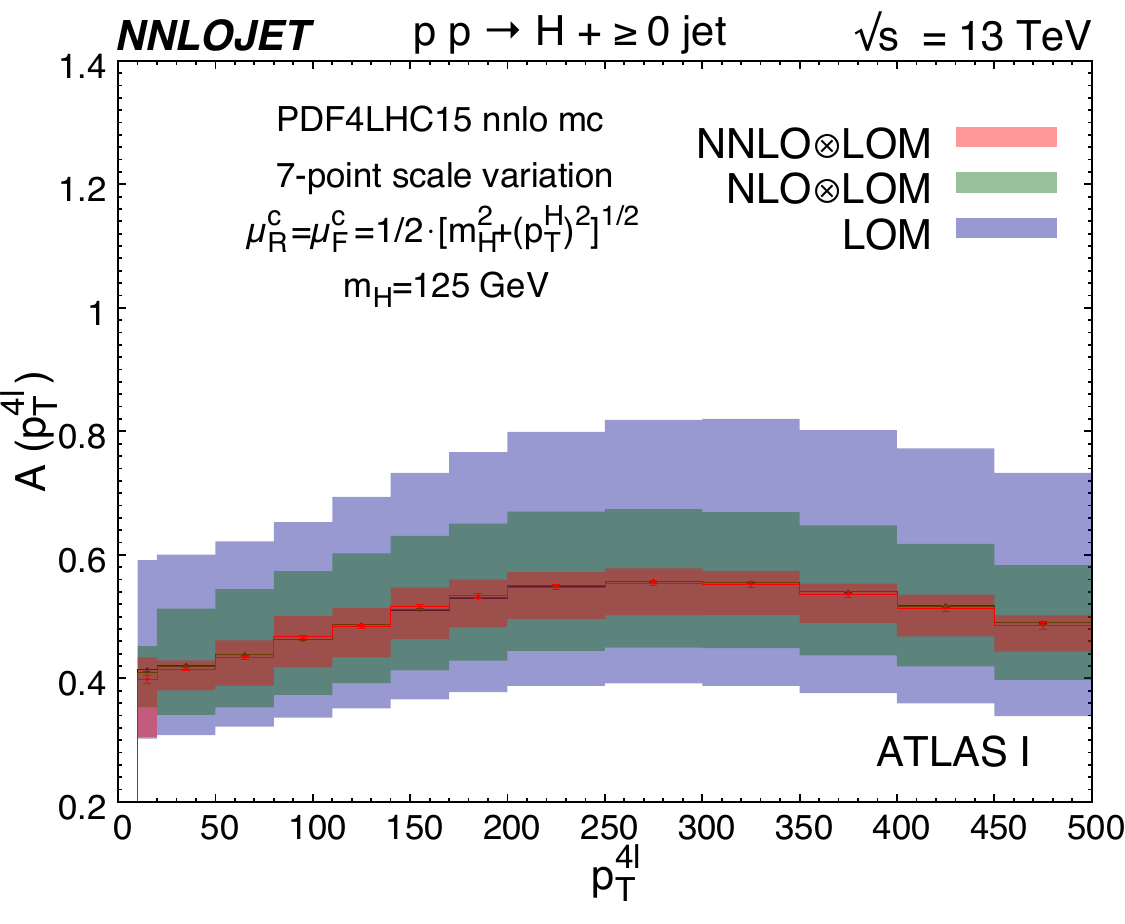}
        \includegraphics[width=0.32\textwidth]{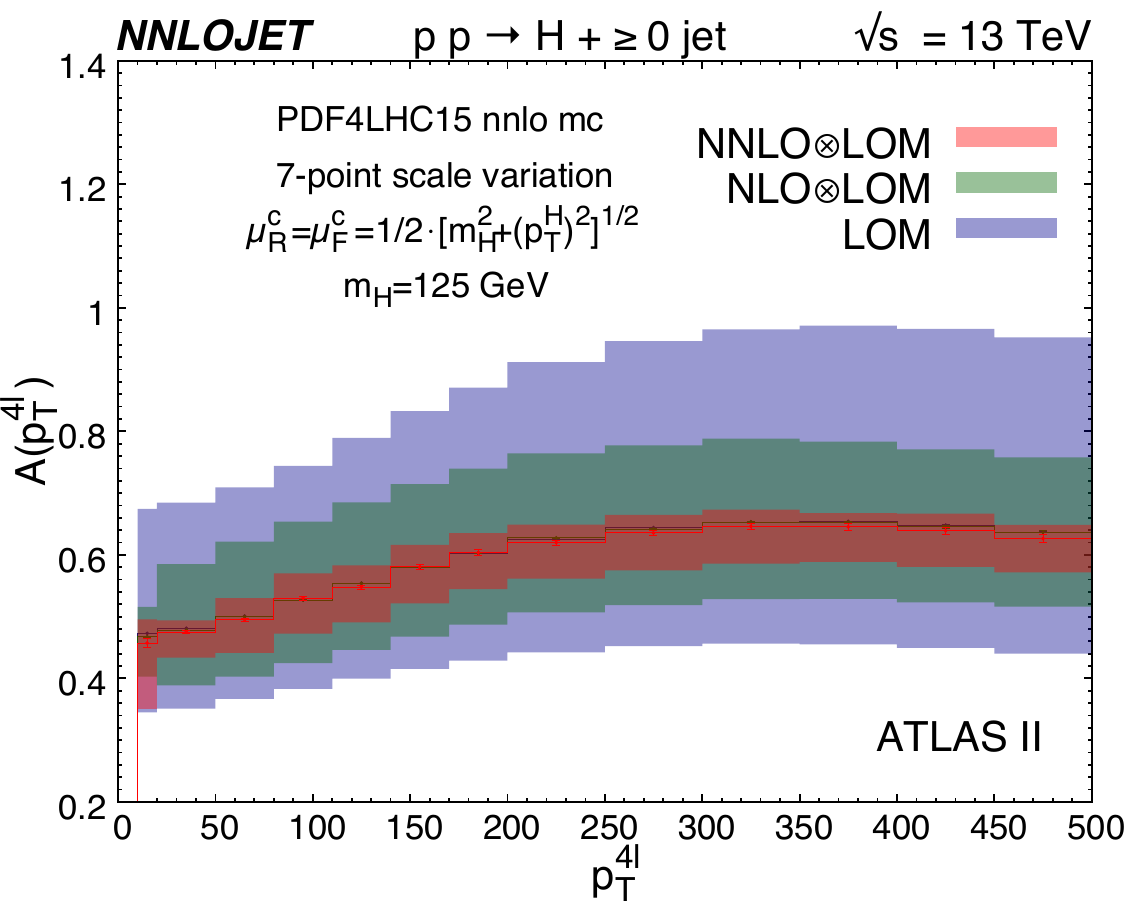}
        \label{fig:acceptHpTandHpTJ-CMS-HpT}
    \end{subfigure}

    \begin{subfigure}[b]{\textwidth}
\centering
        \includegraphics[width=0.32\textwidth]{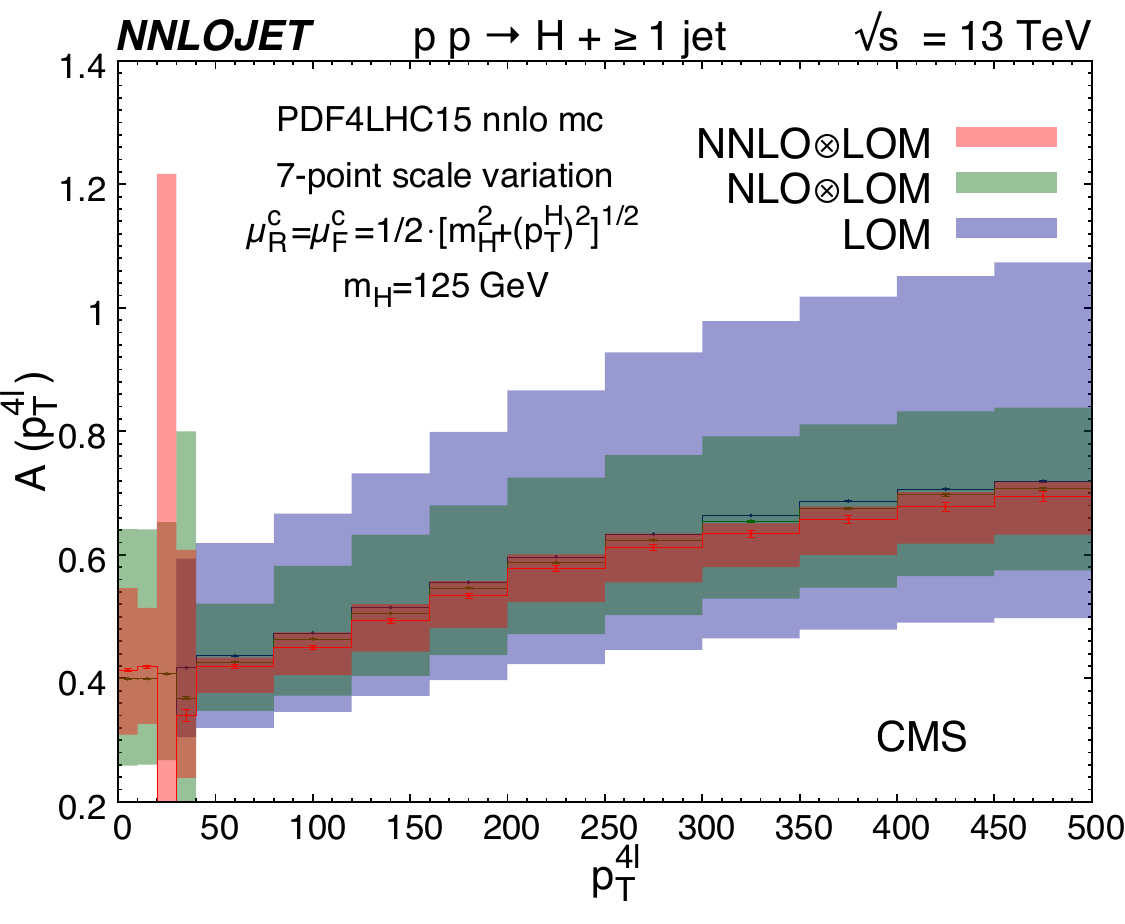}
        \includegraphics[width=0.32\textwidth]{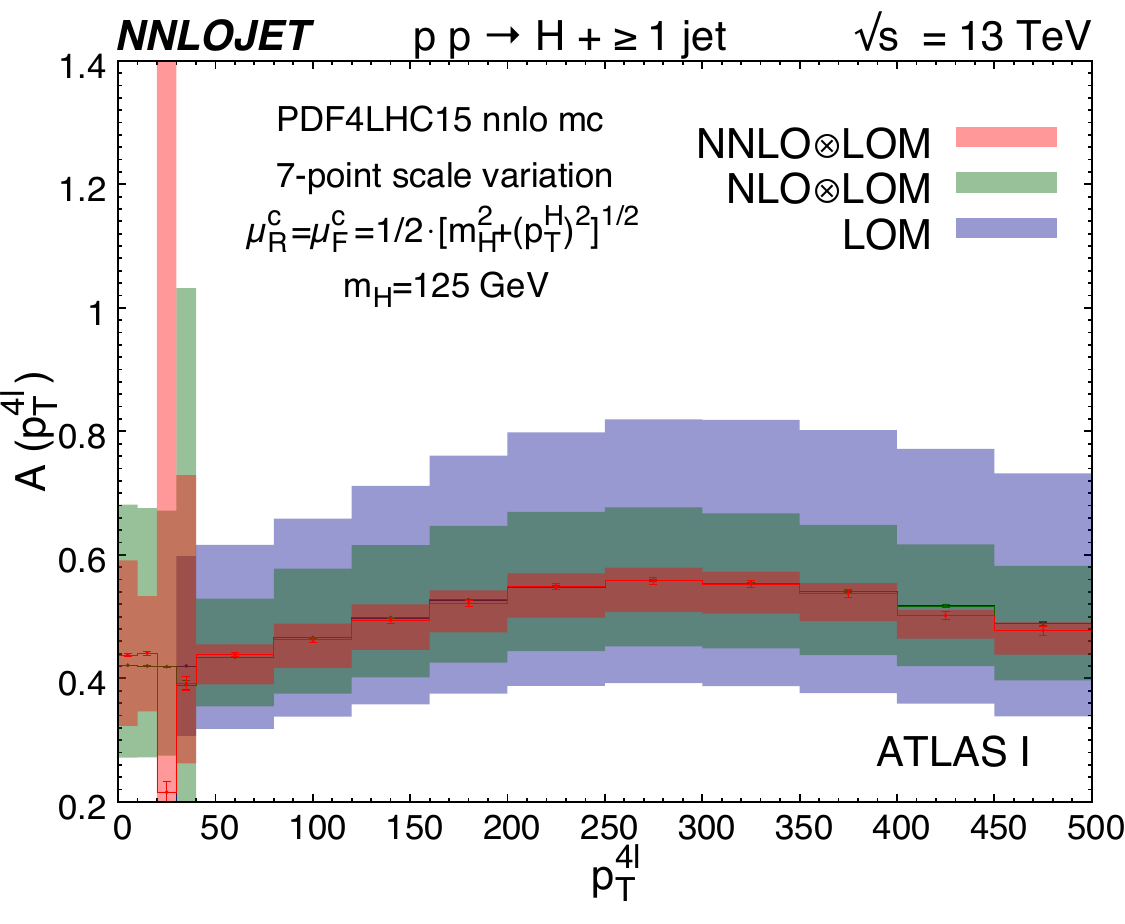} 
        \includegraphics[width=0.32\textwidth]{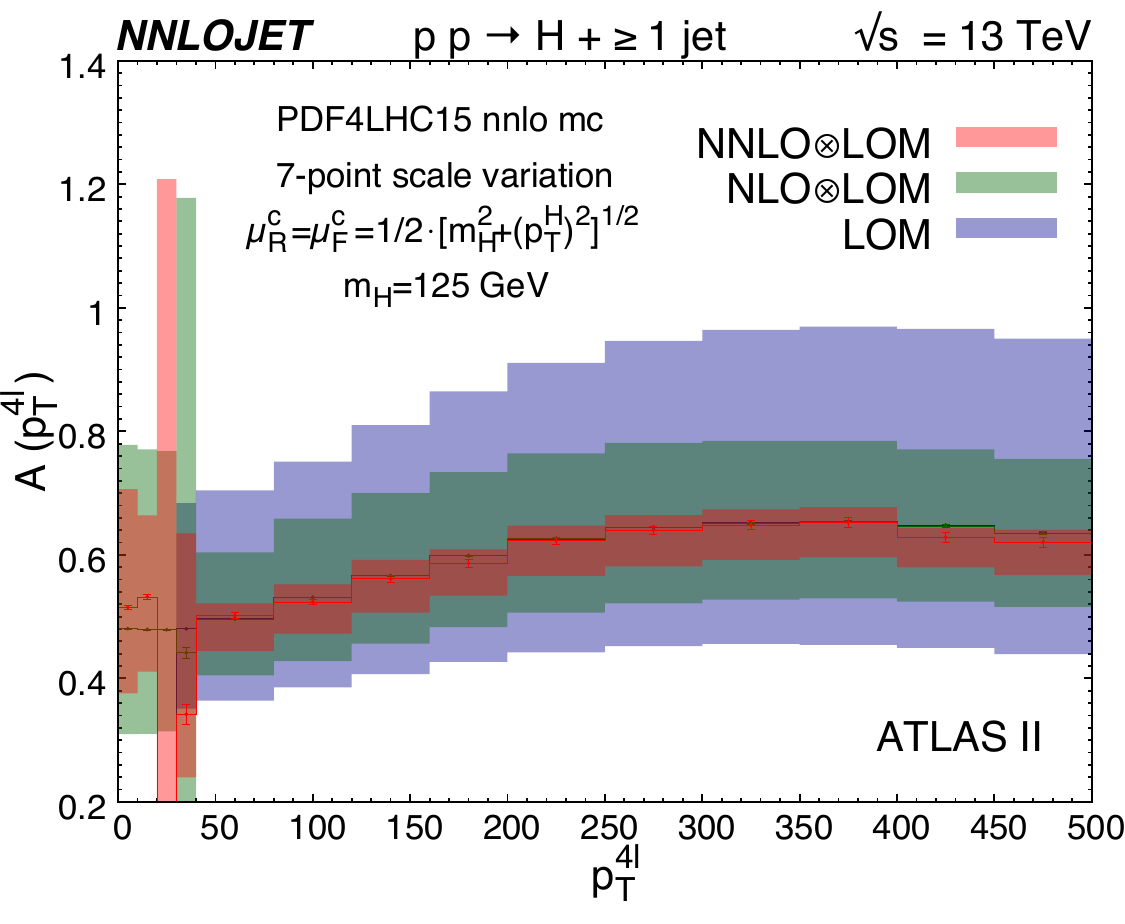}                 
        \label{fig:acceptHpTandHpTJ-ATLAS1-HpT}
    \end{subfigure}
    \caption{Acceptance of inclusive Higgs boson transverse momentum distributions for Higgs (upper) and Higgs-plus-jet (lower) production up to NNLO for CMS (left), \ATLASI (centre) and \ATLASII (right) cuts.}\label{fig:acceptHpTandHpTJ}
\end{figure}

The differential acceptance factors as defined in Eq.~\eqref{eq:accept} for the four-lepton channel are shown as a function of the Higgs boson transverse momentum in 
Fig.~\ref{fig:acceptHpTandHpTJ} for both Higgs and Higgs-plus-jet final states. 
To estimate the scale uncertainties of acceptance factors, as in Fig.~\ref{fig:KfactorHpT}, the uncertainty bands are obtained by fixing the semi-inclusive distributions (Eq.~\eqref{eq:seminc}) in the denominator at the respective central scale while varying the numerator using a seven-point scale variation.
Combined integration errors originated from both inclusive and fiducial transverse momentum distributions are plotted as error bars.
The fiducial transverse momentum distributions for the four-lepton system 
constitute the numerators of the acceptances, see \eqref{eq:accept}. These are the same distributions used as the numerators of the $K$-factors \eqref{eq:NNLOtoLOK}. 
Consequently, the properties of the $K$-factors of Fig.~\ref{fig:KfactorHpT} propagate into Fig.~\ref{fig:acceptHpTandHpTJ} for various fiducial cuts.
The central value of the acceptance distributions are almost identical between LO, NLO and NNLO for  the \ATLASI and \ATLASII analyses. 
Because of the different lepton isolation prescription, in the CMS analysis, the central value of the acceptance for inclusive Higgs boson production and the Higgs-plus-jet production decrease by up to $3.1\%$ from LO to NNLO.
This indicates that the estimation of acceptances from LO simulations do not fully capture the higher-order effects from the lepton isolation prescription, which may in turn affect the 
determination of simplified template cross sections from the four-lepton channel.  
The shapes of the Higgs boson transverse momentum acceptance distributions are in general similar between LO, NLO and NNLO for a fixed set of fiducial cuts. 
Between the different sets of fiducial cuts, we observe a considerable change in the shapes of the acceptance factors, especially in the 
region of large transverse momenta. 
These differences can be explained through the lepton kinematic cuts in the different analyses, as discussed in Fig.~\ref{fig:LOshape} above.
For Higgs-plus-jet acceptance factors, we moreover observe a distortion around  $p_T^{4l}=30$~GeV. This Sudakov shoulder is caused by the requirement of observing at least one jet with 
   $p_T^{j}>30$~GeV, and is well-understood. 
Away from this region, the shape of the $p_T^{4l}$ acceptance distributions are similar for Higgs and Higgs-plus-jet production.   
When including higher order corrections, the jet-lepton cone separation criterion $\Delta R(j,l)$ applied by ATLAS analyses does not cause noticeable deviations 
for the $p_T^{4l}$ acceptance between LO, NLO and NNLO distributions. 
This is partially because of the small cone size required for jet-lepton separation which applies only to relatively limited fiducial volume. 
Moreover, it should be recalled that failing the $\Delta R(j,l)$ requirement will not cause the event to be removed; instead, the number of jets is reduced. 

\begin{figure}
    \centering
    \begin{subfigure}[b]{0.32\textwidth}
        \includegraphics[width=\textwidth]{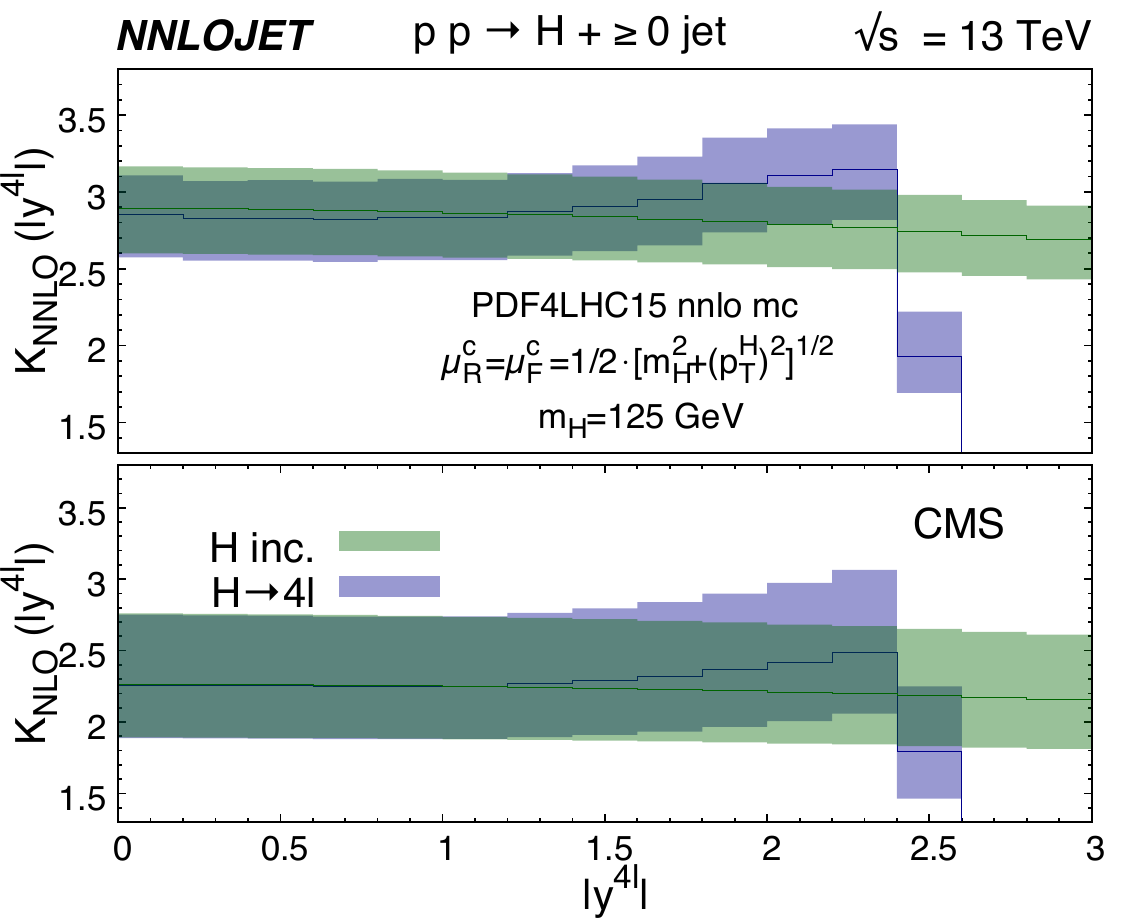}
        \label{fig:CMS-ynorm}
    \end{subfigure}
    \begin{subfigure}[b]{0.32\textwidth}
        \includegraphics[width=\textwidth]{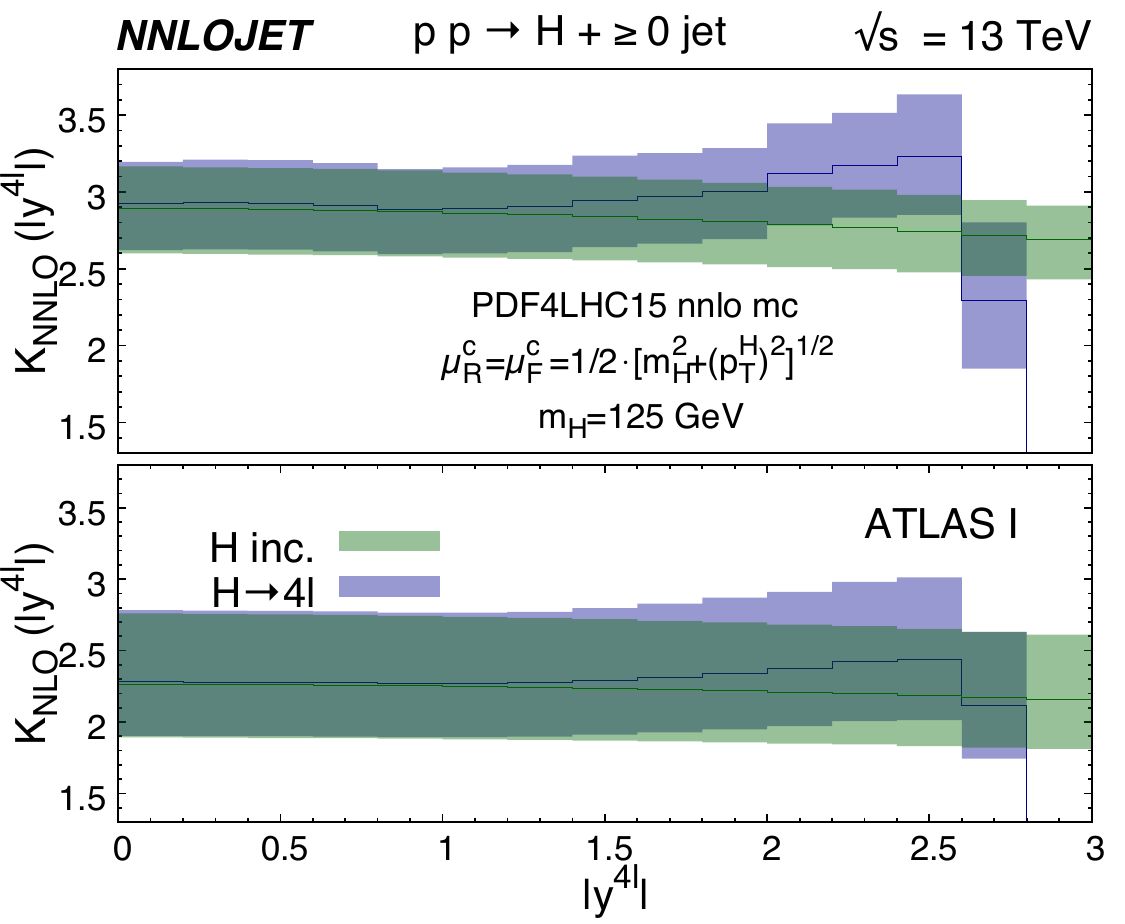}
        \label{fig:ATLAS1-ynorm}
    \end{subfigure}
    \begin{subfigure}[b]{0.32\textwidth}
        \includegraphics[width=\textwidth]{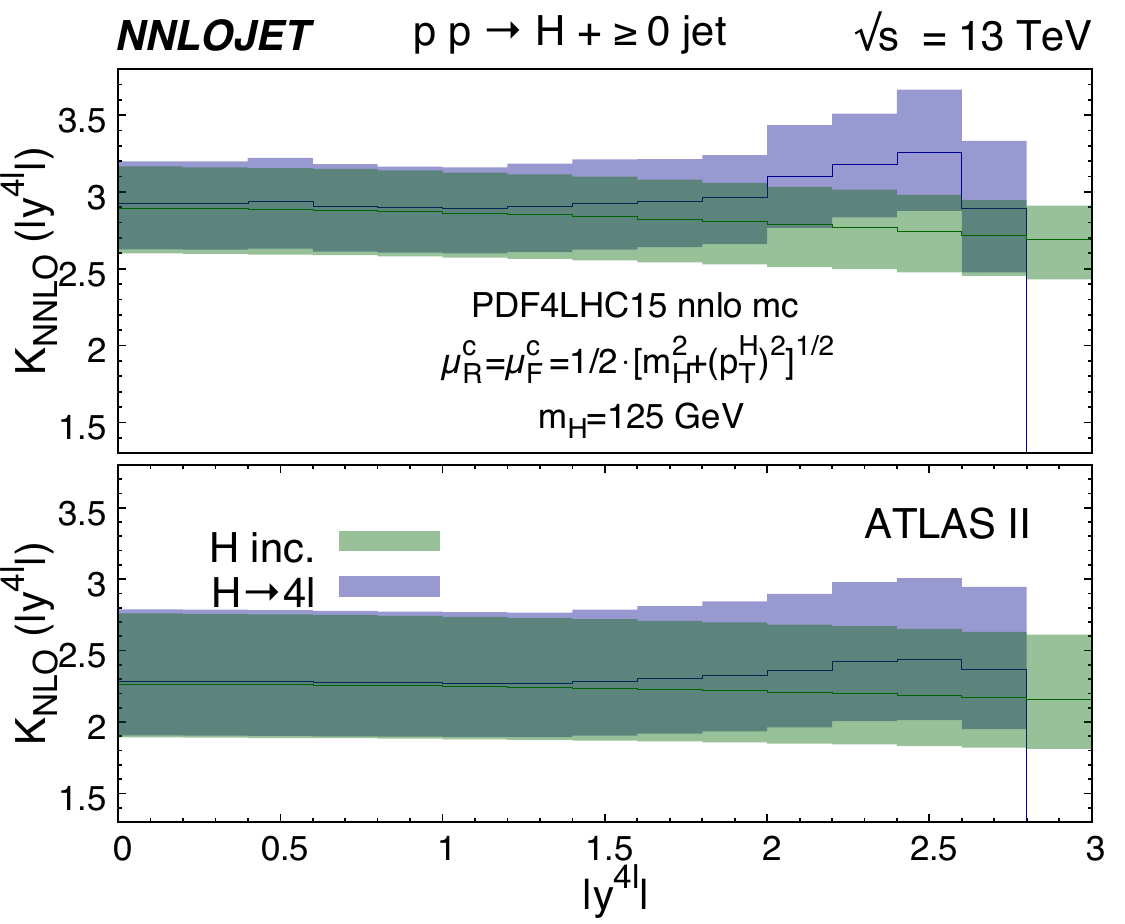}
        \label{fig:ATLAS2-ynorm}
    \end{subfigure}
    \caption{Rapidity distributions of the inclusive Higgs boson or four-lepton system from Higgs boson decay for CMS (left), \ATLASI (centre) and \ATLASII (right) cuts and normalised by the corresponding LO distributions.}\label{fig:Kfactoryh}
\end{figure}

As for the transverse momentum distribution, the Higgs boson rapidity distribution is also inferred from fiducial distributions in specific Higgs boson decay channels. Since the 
rapidity distribution is fully inclusive in the transverse momentum, its perturbative expansion is related to the Higgs-plus-zero-jet process, i.e.\ starting one order lower in 
$\alpha_s$ than the transverse momentum distribution. 
In Fig.~\ref{fig:Kfactoryh}, we compute the $K$-factors for the Higgs boson rapidity distributions up to NNLO for inclusive (green) and fiducial (blue) event selection cuts. 
In the central rapidity region (where the bulk of the cross section is contained), we observe an excellent agreement of inclusive and fiducial $K$-factors 
for the ATLAS cuts, and a slightly lower fiducial $K$-factor for the CMS cuts. This pattern reflects the different lepton isolation prescriptions and was discussed in detail above. 
At larger rapidities, $|y^{4l}|>1.5$, we first observe an increase in the fiducial $K$-factors, which is then followed by a sharp drop that occurs in the region where the rapidity cuts
on the individual leptons become 
effective, and where the cross sections drop to zero. The initial increase in the fiducial $K$-factors can be understood to be due to the extra radiation at higher orders, which, for a given Higgs boson 
rapidity, allows the leptons to be more central, thereby passing the fiducial cuts.

\begin{figure}
    \centering
    \begin{subfigure}[b]{0.32\textwidth}
        \includegraphics[width=\textwidth]{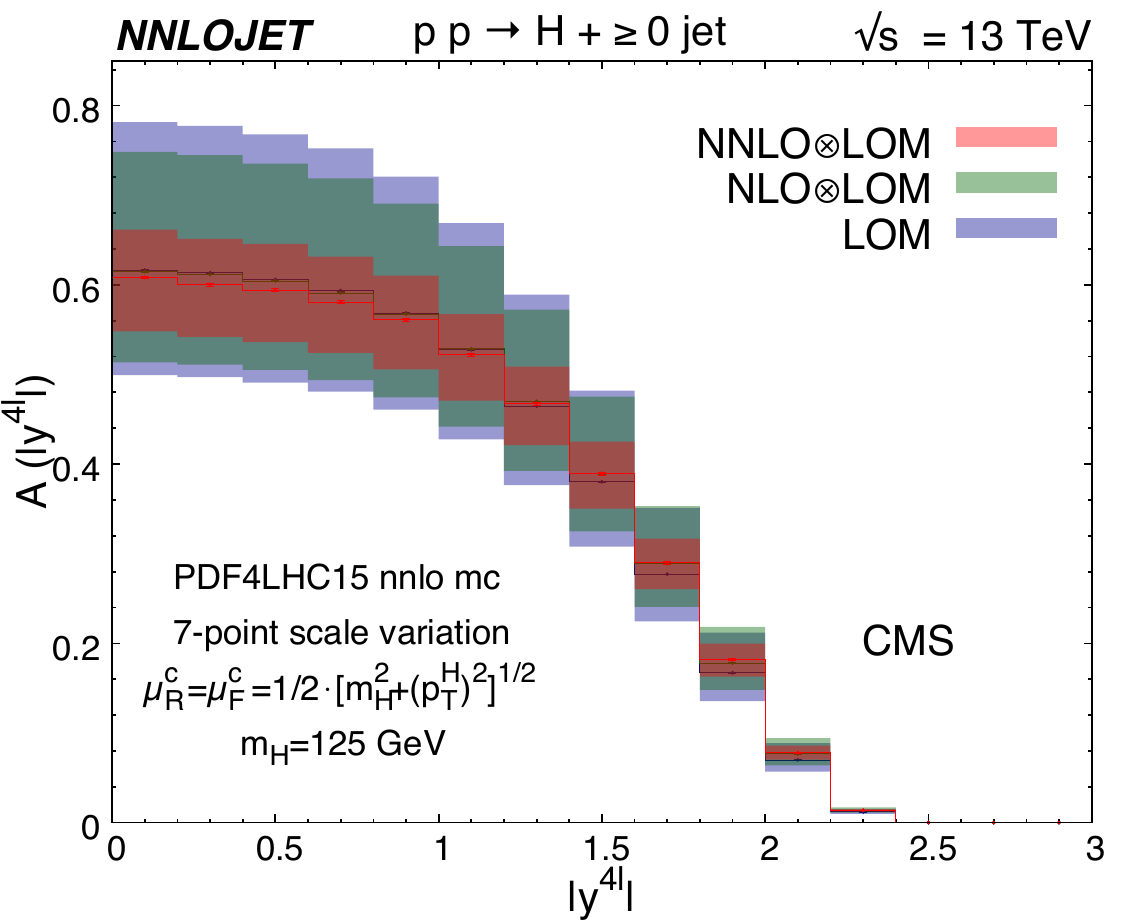}
        \label{fig:CMS-ptratio}
    \end{subfigure}
    \begin{subfigure}[b]{0.32\textwidth}
        \includegraphics[width=\textwidth]{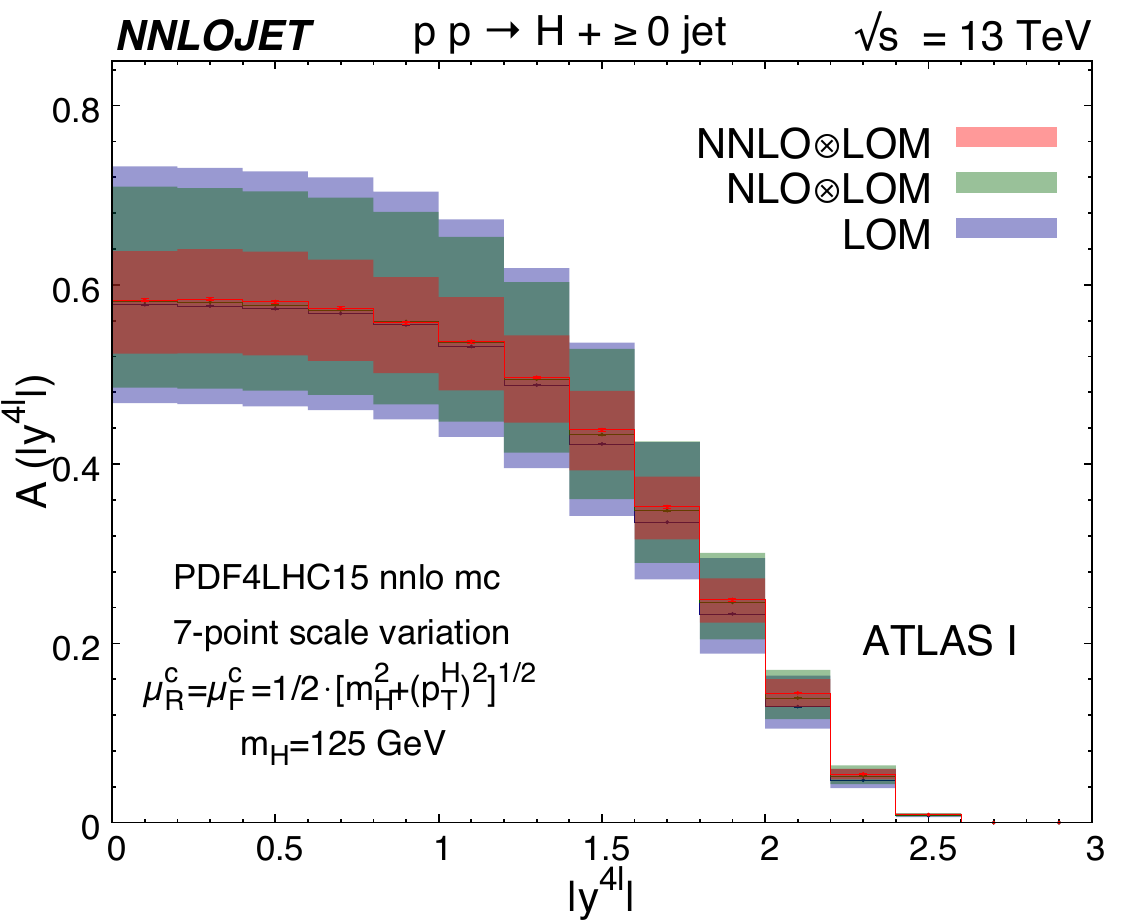}
        \label{fig:ATLAS1-ptratio}
    \end{subfigure}
    \begin{subfigure}[b]{0.32\textwidth}
        \includegraphics[width=\textwidth]{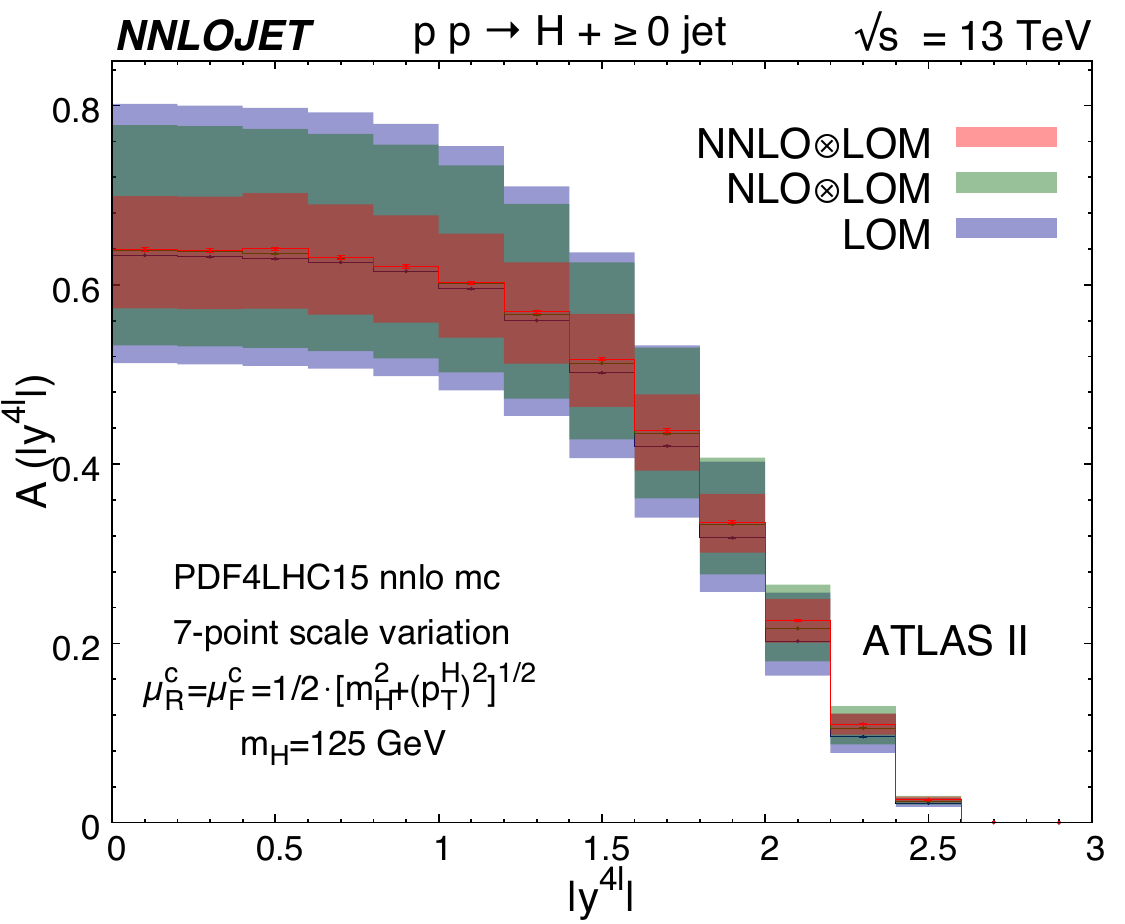}
        \label{fig:ATLAS2-ptratio}
    \end{subfigure}
    \caption{Acceptance of Higgs boson rapidity distributions for Higgs boson production up to NNLO for CMS (left), \ATLASI (centre) and \ATLASII (right) cuts.}\label{fig:acceptyH}
\end{figure}

Figure~\ref{fig:acceptyH} presents the acceptance distributions for the rapidity of the four-lepton system from Higgs boson decay. 
As in Fig.~\ref{fig:Kfactoryh}, in the central rapidity region the shapes of the distributions are less sensitive than the transverse momentum acceptances with respect to different lepton isolation criteria. 
At larger rapidities of $1.5<|y^{4l}|<2.5$, the acceptance distributions become more sensitive  to the different fiducial cuts on the lepton rapidity (which are similar for CMS and \ATLASI, but more loose for \ATLASII).

\subsection{Comparison with LHC measurements}
\label{sec:comparison}
The ATLAS and CMS experiments have measured fiducial cross sections in the four-lepton decay mode for Higgs and Higgs-plus-jet production at 13~TeV. A summary of the fiducial cuts 
applied in the ATLAS~\cite{ATLASHto4l,ATLASHto4lconf} and CMS~\cite{CMSHto4l,CMSHto4lconf} measurements are summarised in Table~\ref{tab:allthree}. Both CMS 
measurements use the same fiducial cuts. In the following, we compare these measurements to the
predictions obtained using \NNLOJET for the gluon-fusion production to NNLO in the infinite-top-mass EFT, re-weighted for quark mass effects by the exact LO expressions, as described in 
Section~\ref{subsec:quarkmass} above. 
The Higgs boson decay to four leptons is described by using the narrow-width approximation with an on-shell Higgs boson decaying to two off-shell Z bosons, each of which decay to a lepton-antilepton pair. 
\begin{figure}
    \centering
    \begin{subfigure}[b]{0.32\textwidth}
        \includegraphics[width=\textwidth]{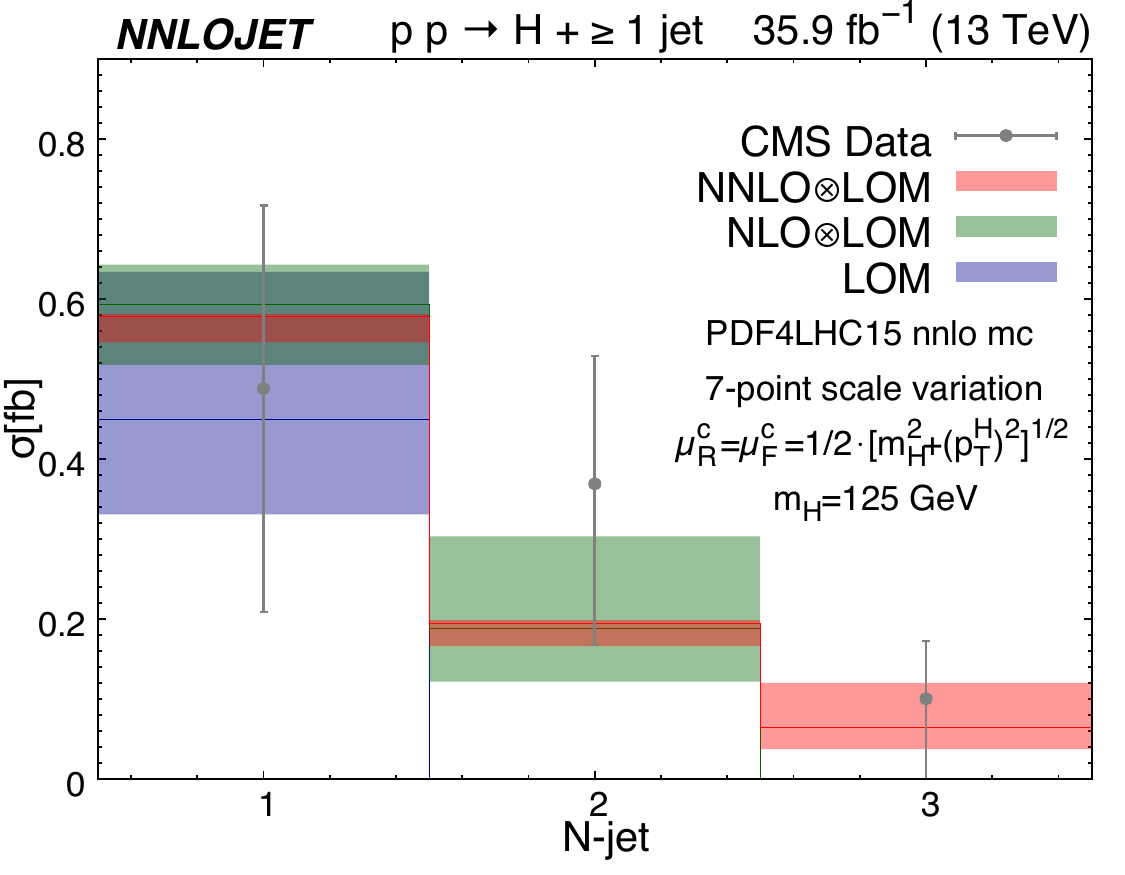}
    \end{subfigure}
    \begin{subfigure}[b]{0.32\textwidth}
        \includegraphics[width=\textwidth]{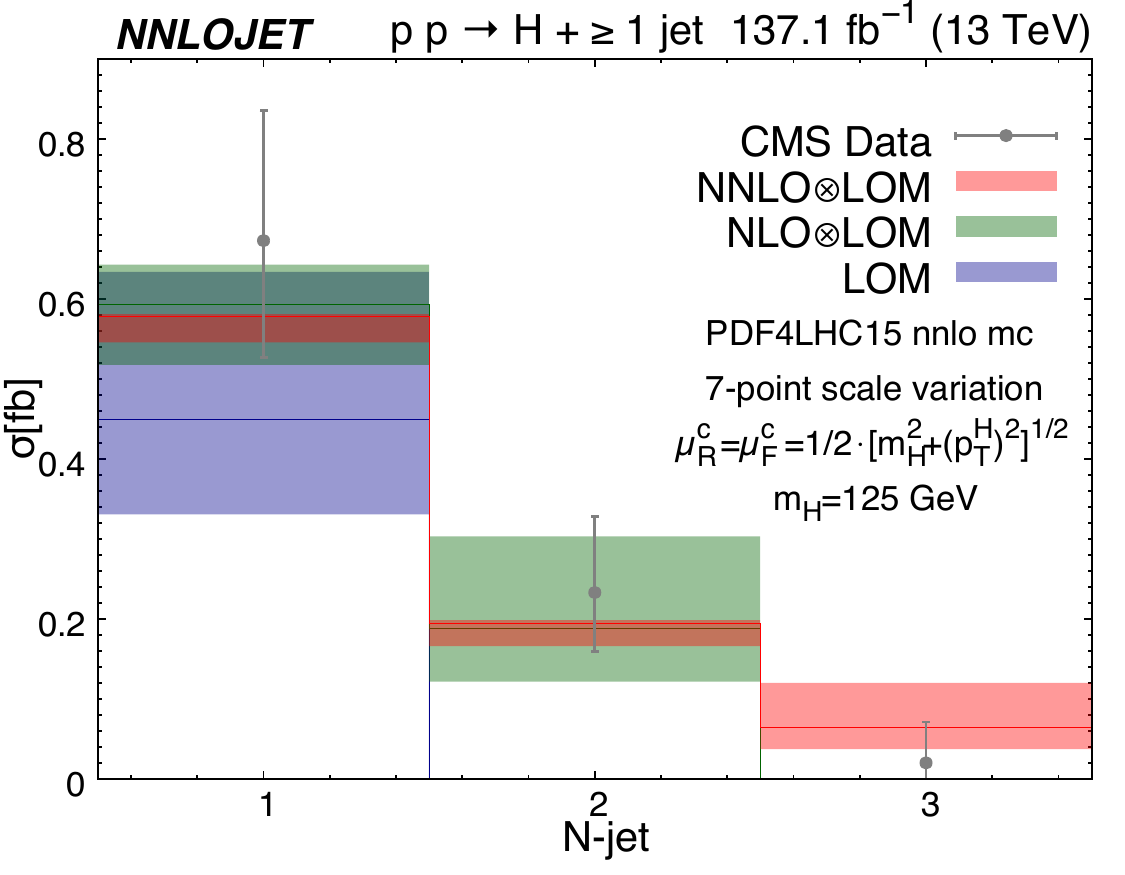}
    \end{subfigure}

    \begin{subfigure}[b]{0.32\textwidth}
        \includegraphics[width=\textwidth]{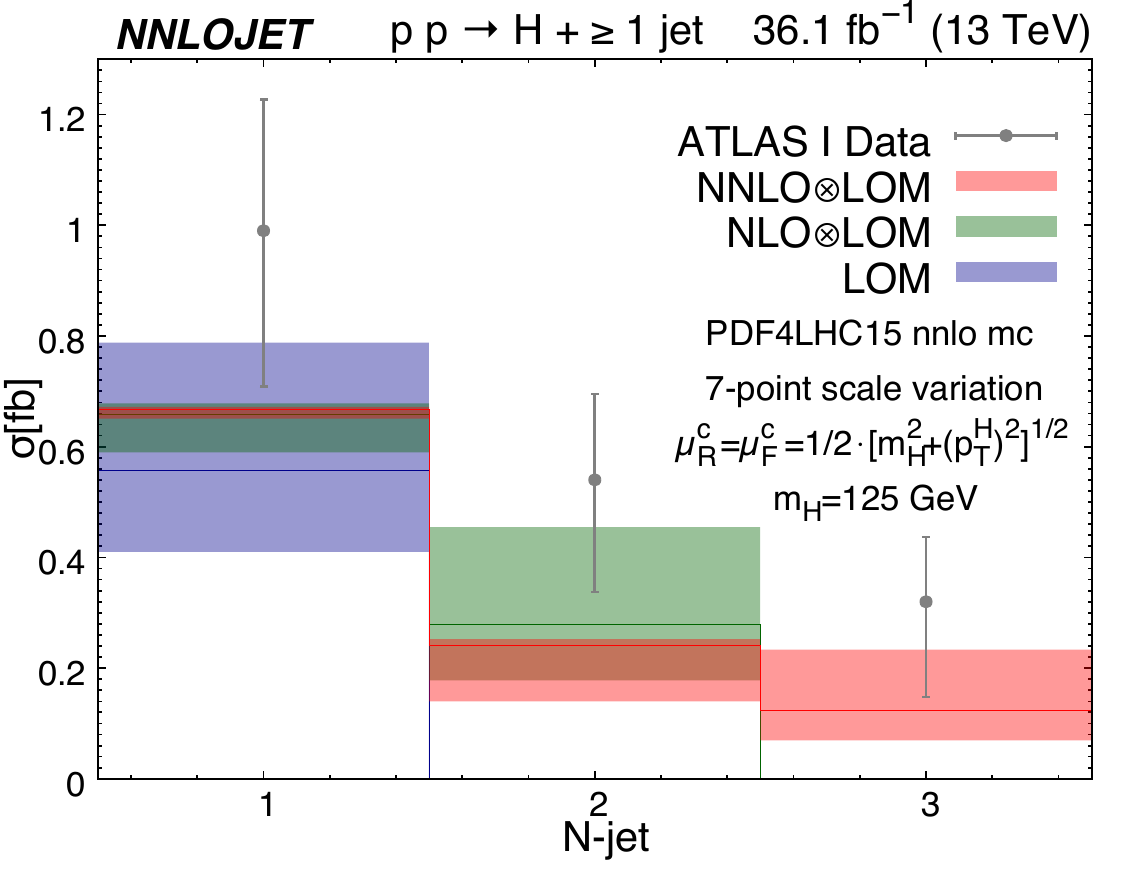}
    \end{subfigure}
    \begin{subfigure}[b]{0.32\textwidth}
        \includegraphics[width=\textwidth]{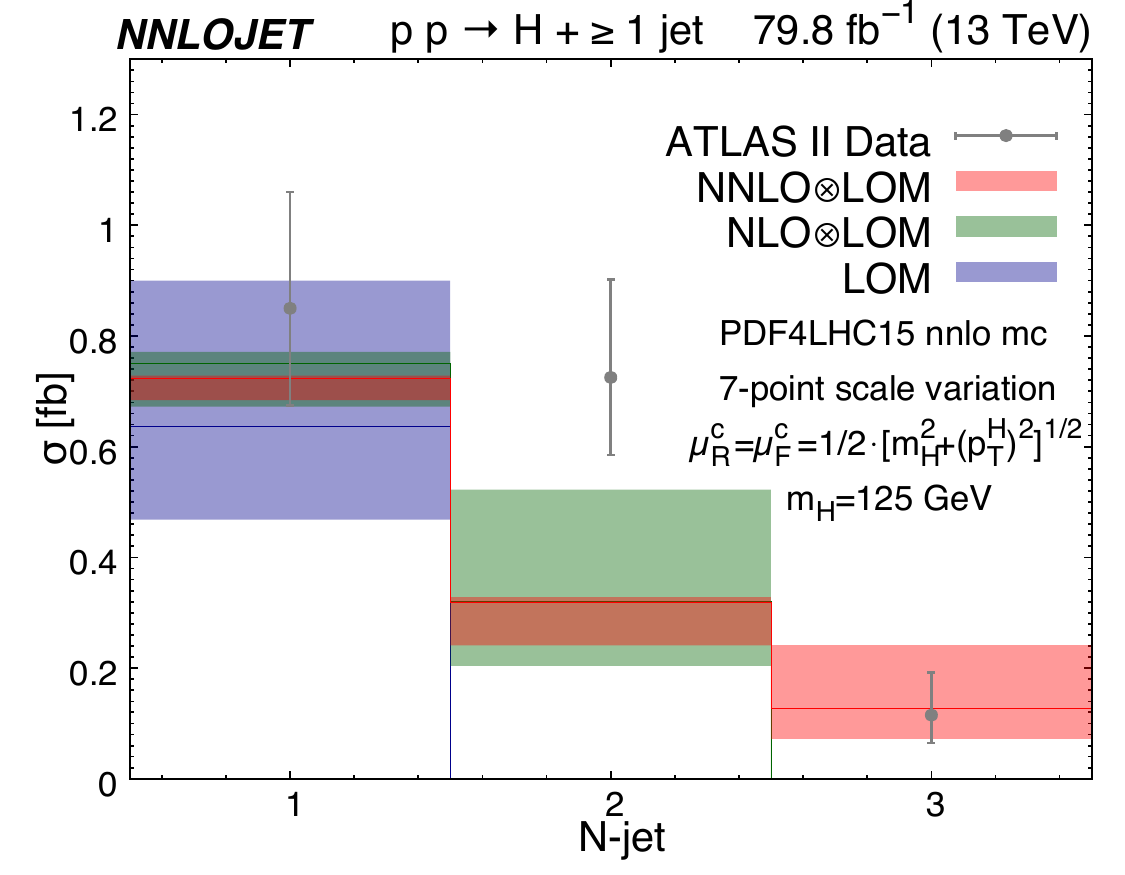}
    \end{subfigure}
    \caption{Jet multiplicity in Higgs-plus-jet production up to NNLO for CMS [$\int {\cal L}dt = $35.9~fb$^{-1}$] (upper-left) and [$\int {\cal L}dt = $137.1~fb$^{-1}$] (upper-right), \ATLASI [$\int {\cal L}dt = $36.1~fb$^{-1}$] (lower-left) and \ATLASII [$\int {\cal L}dt = $79.8~fb$^{-1}$] (lower-right) cuts and integrated luminosities. NLO and NNLO EFT predictions are re-weighted by LO quark mass effects.}\label{fig:njets}
\end{figure}

Figure~\ref{fig:njets} compares the ATLAS and CMS measurements for different jet multiplicities with the \NNLOJET predictions up to $\mathcal{O}(\alpha_s^5)$, which is up to NNLO accuracy for Higgs-plus-one-jet final states (but only LO for Higgs-plus-three-jets). 
We observe a significant reduction of scale uncertainties for the one-jet bin going from NLO to NNLO for all three sets of fiducial cuts. 
The more recent \ATLASII~\cite{ATLASHto4lconf} (79.8~fb$^{-1}$) and CMS~\cite{CMSHto4lconf}  (137.1~fb$^{-1}$) measurements of the Higgs-plus-one-jet cross section  agree well with NNLO theory
predictions. Higgs-plus-two-jet production is described only to NLO, showing good agreement with CMS~\cite{CMSHto4lconf}, while the \ATLASII~\cite{ATLASHto4lconf} measurement is considerably 
above the theory prediction. 
\begin{table}
\begin{center}
\begin{tabular}{lccc}
\hline
\hline
 Cross section [fb]& \ATLASII Data~\cite{ATLASHto4lconf}   & NNLO  & NNLO$\otimes$LOM  \\
\hline
$\sigma_{4\mu}$       & $0.97\pm 0.17 \pm0.05$  & $0.700^{+0.066}_{-0.072}$ & $0.660^{+0.062}_{-0.068}$ \\
$\sigma_{4e}$         & $0.61\pm 0.21 \pm0.07$  & $0.700^{+0.066}_{-0.072}$ & $0.660^{+0.062}_{-0.068}$ \\
$\sigma_{2\mu 2e}$    & $0.88\pm 0.21 \pm0.08$  & $0.617^{+0.058}_{-0.064}$ & $0.582^{+0.055}_{-0.060}$ \\
$\sigma_{2e2\mu}$     & $1.37\pm 0.22 \pm0.07$  & $0.617^{+0.058}_{-0.064}$ & $0.582^{+0.055}_{-0.060}$ \\
\hline
$\sigma_{4l}$ & $3.84\pm 0.41 \pm0.23$  & $2.65^{+0.250}_{-0.272}$  & $2.48^{+0.234}_{-0.255}$ \\
\hline
\hline
$\sigma_{H}$ [pb]     & $67.2 \pm 6.8 \pm 4.1$  & $42.5^{+3.86}_{-4.28}$  & $40.1^{+3.64}_{-4.03}$ \\
\hline
\hline
\end{tabular}
\end{center}
\caption{The fiducial cross section of Higgs boson production measured in the four-lepton final state using \ATLASII fiducial cuts listed in Table~\ref{tab:allthree}. The decay channel of $2e2\mu$ is distinguished from $2\mu2e$ by requiring the first lepton pair being the $Z_1$ candidate as described in Section~\ref{sec:fiducial}. Experimental errors are statistical and systematic. Theoretical uncertainties are from scale variation as described in the text.\label{tab:fidATLASIINNLO}}
\end{table}

A similar discrepancy between \ATLASII data~\cite{ATLASHto4lconf}  and theory can also be observed in the total fiducial cross section given in Table~\ref{tab:fidATLASIINNLO}. 
The measured total cross sections for reconstructed inclusive Higgs boson production ($\sigma_{H}$) and for Higgs boson decay to different lepton flavour combinations ($\sigma_{4\mu},\sigma_{4e},\sigma_{2e2\mu}$ and $\sigma_{2\mu 2e}$) are consistently above the NNLO and NNLO$\otimes$LOM predictions. 
For inclusive Higgs boson production cross sections, sub-dominant production channels together contribute about $15\%$ of the total cross section~\cite{YR4} which can not explain the difference of about $68\%$ between \ATLASII data and NNLO$\otimes$LOM predictions. 

The comparison of the total fiducial cross section between CMS measurement~\cite{CMSHto4lconf} and theory presents better agreement than ATLAS. As summarised in Table~\ref{tab:fidCMSIINNLO}, the NNLO and NNLO$\otimes$LOM predictions are consistently below the CMS data for individual and combined lepton flavour channels, while still being consistent within errors. 
The excess is compatible with the anticipated contribution from 
sub-leading production channels, which will need to be accounted for in the theory predictions once the measurements become more precise with an increasing data set. 


\begin{table}
\begin{center}
\begin{tabular}{lccc}
\hline
\hline
 Cross section [fb]& CMS Data~\cite{CMSHto4lconf}  & NNLO  & NNLO$\otimes$LOM  \\
\hline
$\sigma_{4\mu}$                 & $0.749^{+0.109}_{-0.098}$ & $0.643^{+0.056}_{-0.063}$ & $0.595^{+0.052}_{-0.058}$ \\
$\sigma_{4e}$                   & $0.692^{+0.188}_{-0.160}$ & $0.574^{+0.049}_{-0.056}$ & $0.531^{+0.045}_{-0.052}$ \\
$\sigma_{2\mu 2e + 2e 2\mu}$    & $1.295^{+0.190}_{-0.176}$ & $1.077^{+0.095}_{-0.107}$ & $0.998^{+0.086}_{-0.099}$ \\
\hline
$\sigma_{4l}$                   & $2.733^{+0.330}_{-0.292}$ & $2.294^{+0.200}_{-0.226}$  & $2.124^{+0.186}_{-0.209}$ \\
\hline
\hline
\end{tabular}
\end{center}
\caption{The fiducial cross section of Higgs boson production measured in the four-lepton final state using CMS fiducial cuts listed in Table~\ref{tab:allthree}. The decay channels of $2e2\mu$ and $2\mu2e$ are not distinguished. Experimental uncertainties are statistical and systematic errors combined. Theoretical uncertainties are from scale variation as described in the text.\label{tab:fidCMSIINNLO}}
\end{table}

\begin{figure}
    \centering
    \begin{subfigure}[b]{0.32\textwidth}
        \includegraphics[width=\textwidth]{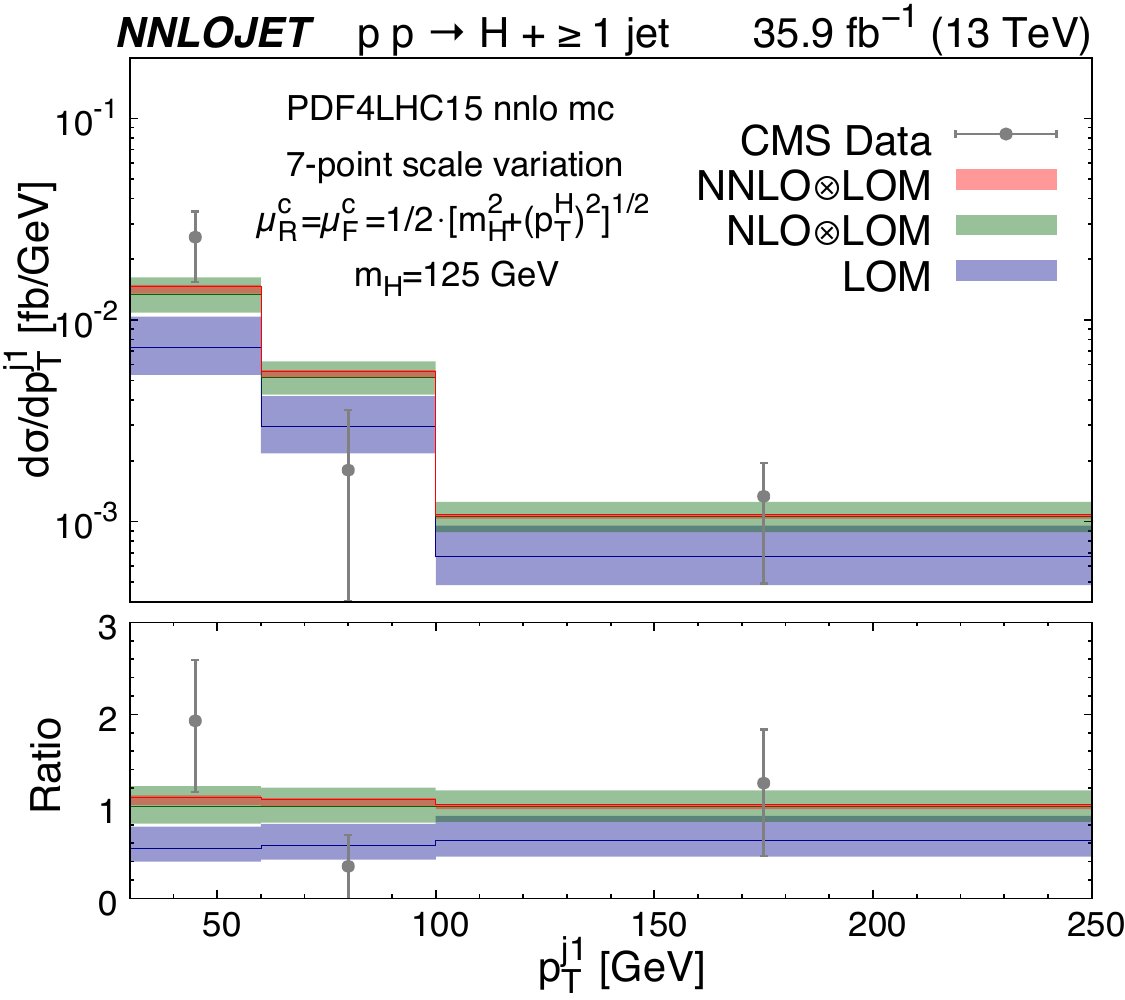}
        \label{fig:CMS-ptj1}
    \end{subfigure}
    \begin{subfigure}[b]{0.32\textwidth}
        \includegraphics[width=\textwidth]{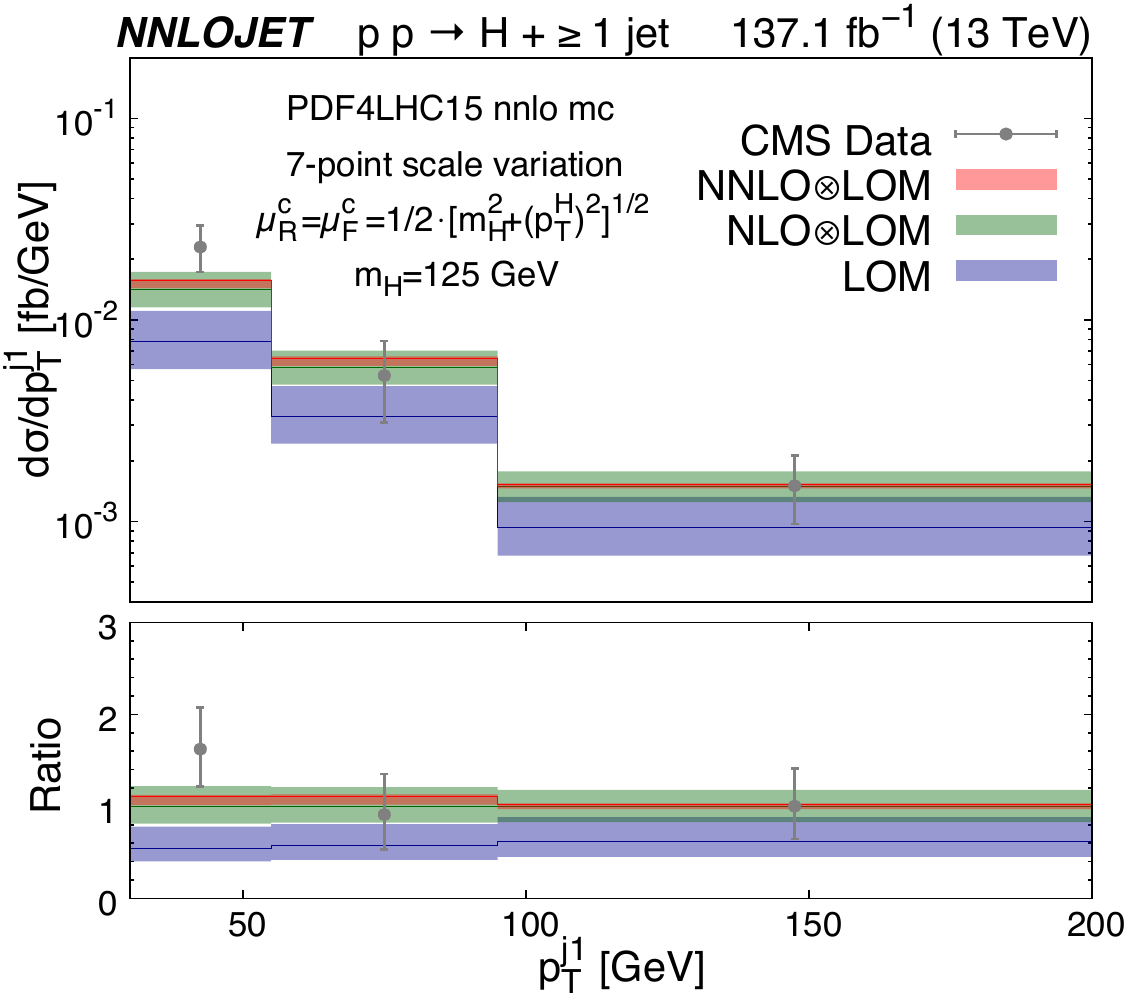}
        \label{fig:CMSII-ptj1}
    \end{subfigure}
    \begin{subfigure}[b]{0.32\textwidth}
        \includegraphics[width=\textwidth]{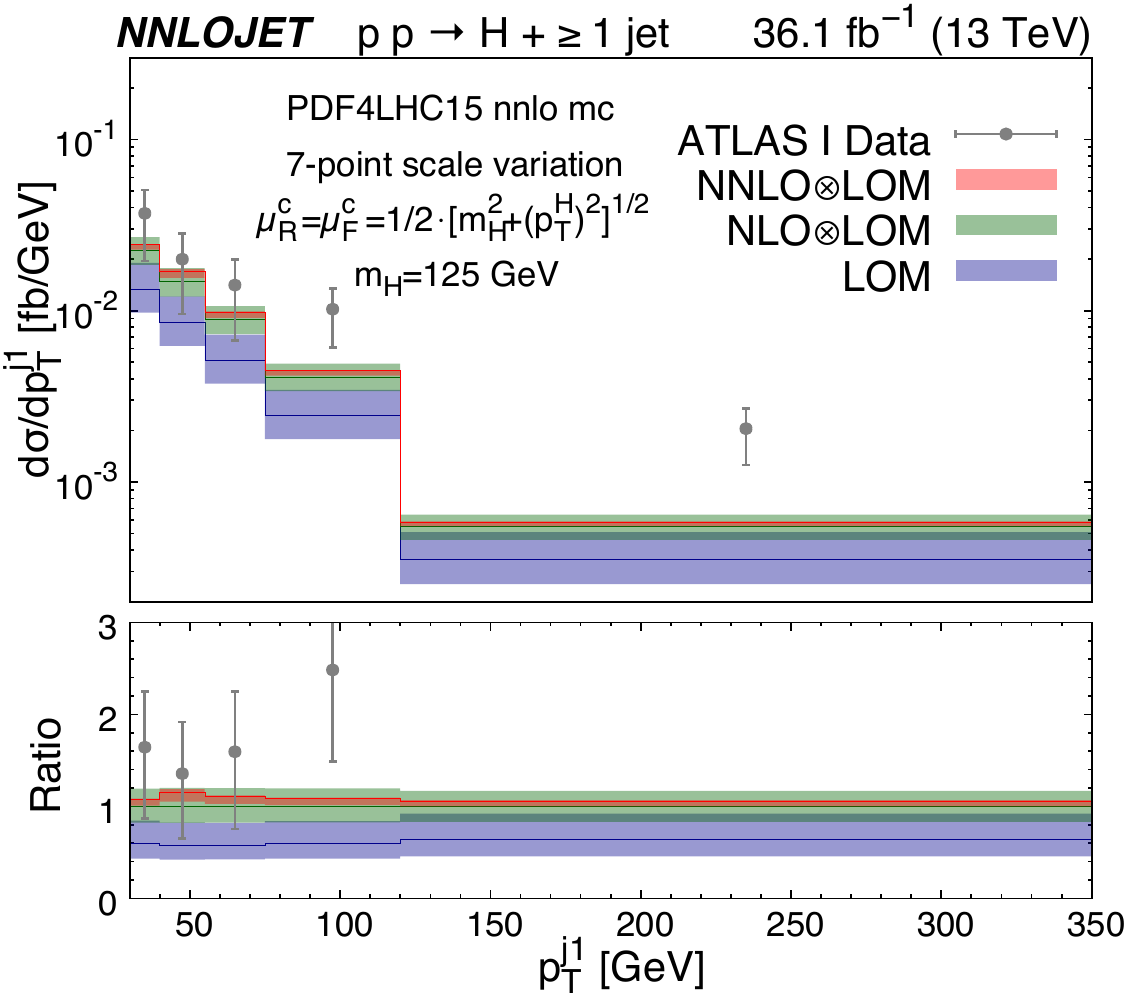}
        \label{fig:ATLAS1-ptj1}
    \end{subfigure}
    \caption{Transverse momentum distributions of the leading jet produced in association 
with a Higgs boson for CMS [$\int {\cal L}dt = $35.9~fb$^{-1}$] (left) and [$\int {\cal L}dt = $137.1~fb$^{-1}$] (centre) and \ATLASI [$\int {\cal L}dt = $36.1~fb$^{-1}$] (right)  cuts and integrated luminosities.}\label{fig:ptj1}
\end{figure}

Both ATLAS~\cite{ATLASHto4l} and CMS~\cite{CMSHto4l,CMSHto4lconf} measured the transverse momentum distributions of the leading jet in Higgs-plus-jet events. 
We compare these measurements to \NNLOJET predictions in Fig.~\ref{fig:ptj1}. 
We observe that the NNLO corrections are sizeable especially in the low transverse momentum region for $p_T^{j1}<100$~GeV. They are slightly larger for the 
\ATLASI cuts than for the CMS cuts, which may be understood from the larger jet rapidity acceptance in the \ATLASI analysis. While observing excellent agreement of the 
theory description with the CMS measurement~\cite{CMSHto4lconf}, we find the \ATLASI data~\cite{ATLASHto4l} to be consistently above the theory prediction, as already observed 
on the jet-binned cross sections in Fig.~\ref{fig:njets}. Unfortunately, a measurement of the jet transverse momentum distribution has not been performed on the larger \ATLASII data set. 

\begin{figure}
    \centering
    \begin{subfigure}[b]{0.32\textwidth}
        \includegraphics[width=\textwidth]{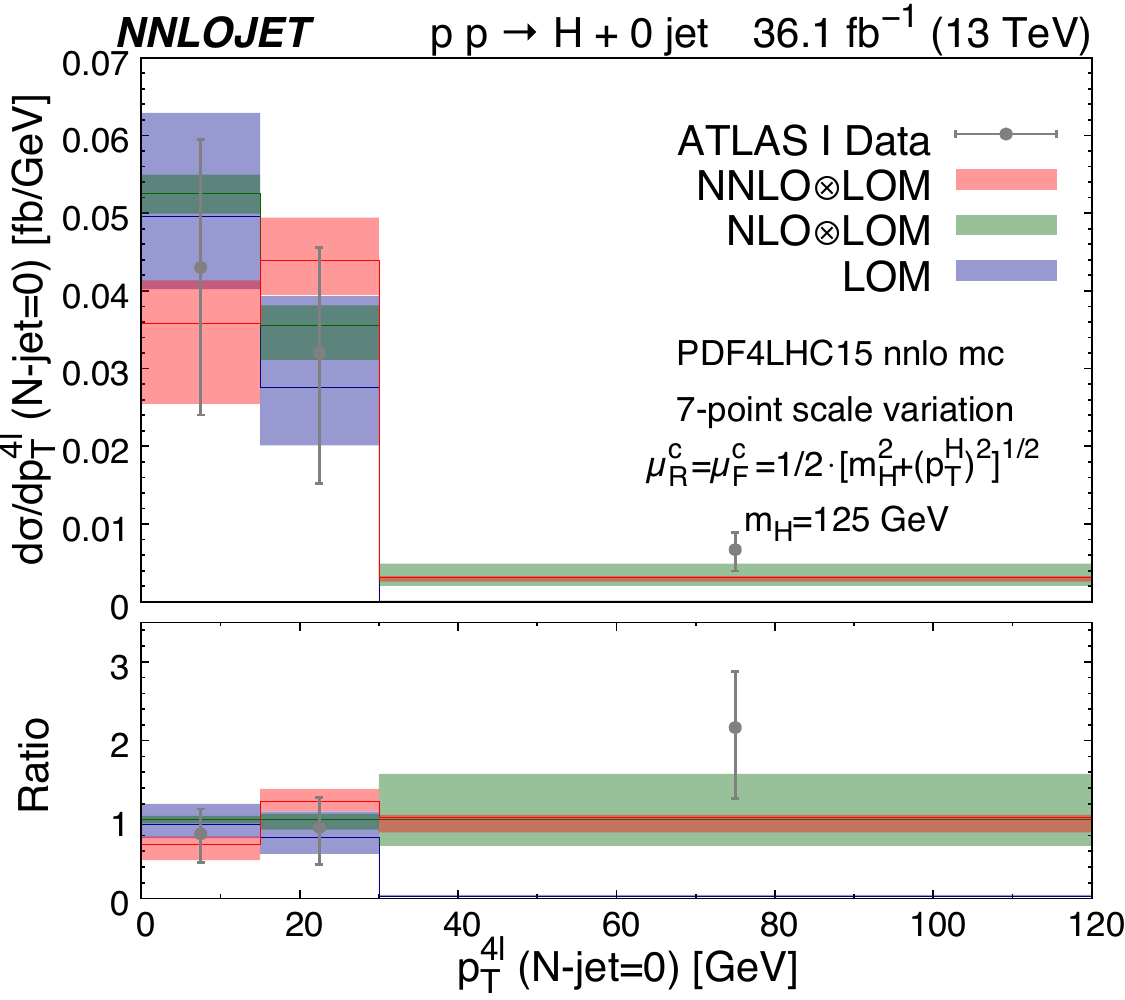}
    \end{subfigure}
    \begin{subfigure}[b]{0.32\textwidth}
        \includegraphics[width=\textwidth]{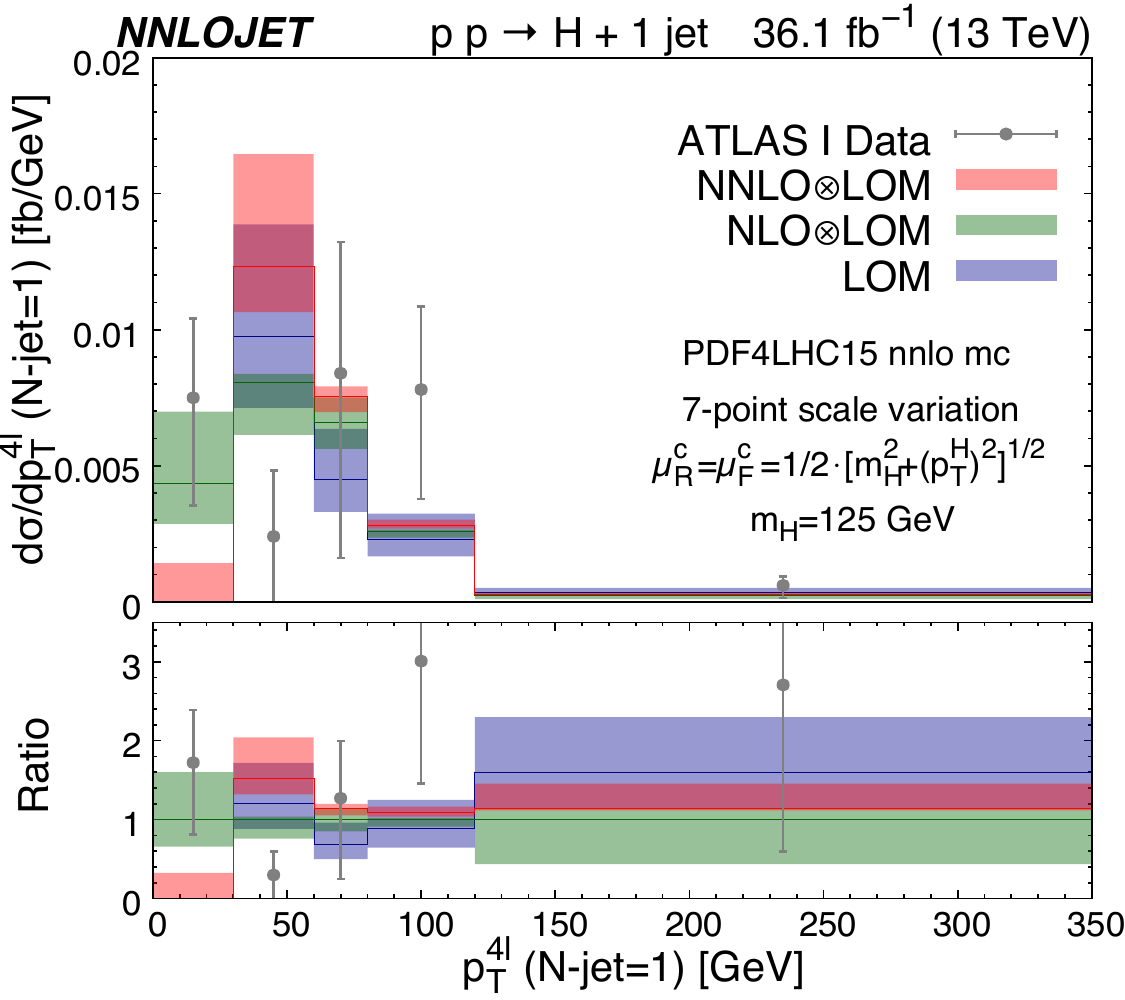}
    \end{subfigure}
    \caption{Transverse momentum distributions of the Higgs boson produced in association with zero (left) and one (right) jet for \ATLASI cuts and integrated luminosity.}\label{fig:pthexclu}
\end{figure}

Figure~\ref{fig:pthexclu} shows the transverse momentum distributions of the Higgs boson produced in association with exactly zero or one jet, which was only measured in the 
\ATLASI analysis~\cite{ATLASHto4l}, with a $p_T^j$ cut of 30~GeV. 
Despite the relatively large experimental errors, it demonstrates the feasibility of determining the Higgs boson transverse momentum distribution in exclusive jet bins, which will be 
instrumental in the future to disentangle different Higgs boson production processes. 
For the Higgs boson transverse momentum distribution with a jet veto (zero-jet exclusive), the first bin is computed from Higgs boson production up to NNLO, while the second and the third bin are obtained 
from Higgs-plus-jet final state up to NNLO. In the one-jet bin, the Higgs boson transverse momentum distribution does not receive a LO contribution below 30~GeV, and consequently 
displays a Sudakov shoulder leading to large corrections and poor perturbative convergence in the first and second bin. At the level of the inclusive Higgs production cross section, the 
effect of this kinematical requirement  can be viewed as a jet veto, whose resummation is well-understood~\cite{jetveto}. To obtain reliable predictions for the Higgs transverse momentum 
distribution in Higgs-plus-jet final states in the vicinity of the $p_T^j$ cut will require these jet veto resummations to be extended towards more exclusive final states, similar to~\cite{jetvetoex}. 
Outside this region, we observe that the NNLO$\otimes$LOM theory predictions provide a good description of the ATLAS data, and that the remaining theory uncertainty is below 10\%. 

\begin{figure}
    \centering
    \begin{subfigure}[b]{0.32\textwidth}
        \includegraphics[width=\textwidth]{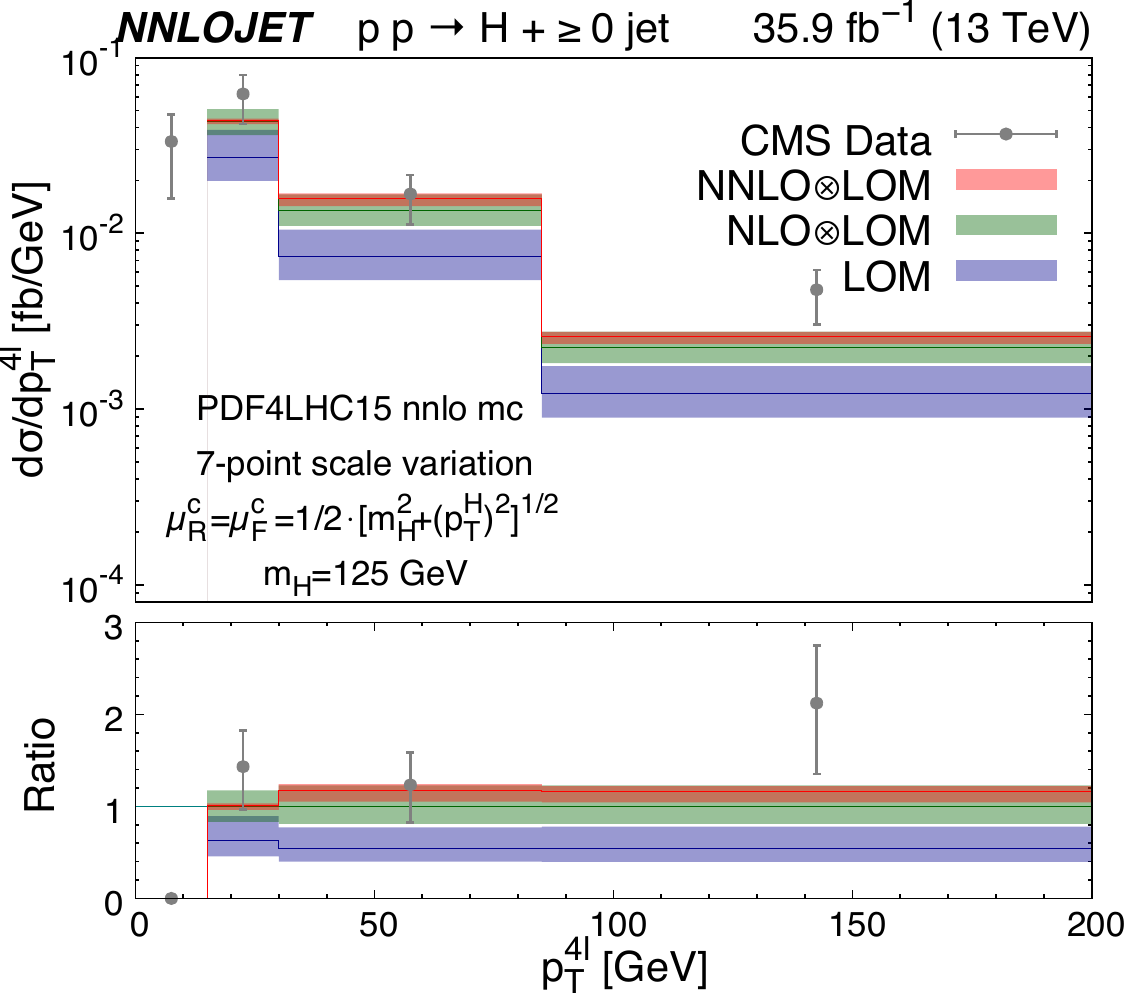}
    \end{subfigure}
    \begin{subfigure}[b]{0.32\textwidth}
        \includegraphics[width=\textwidth]{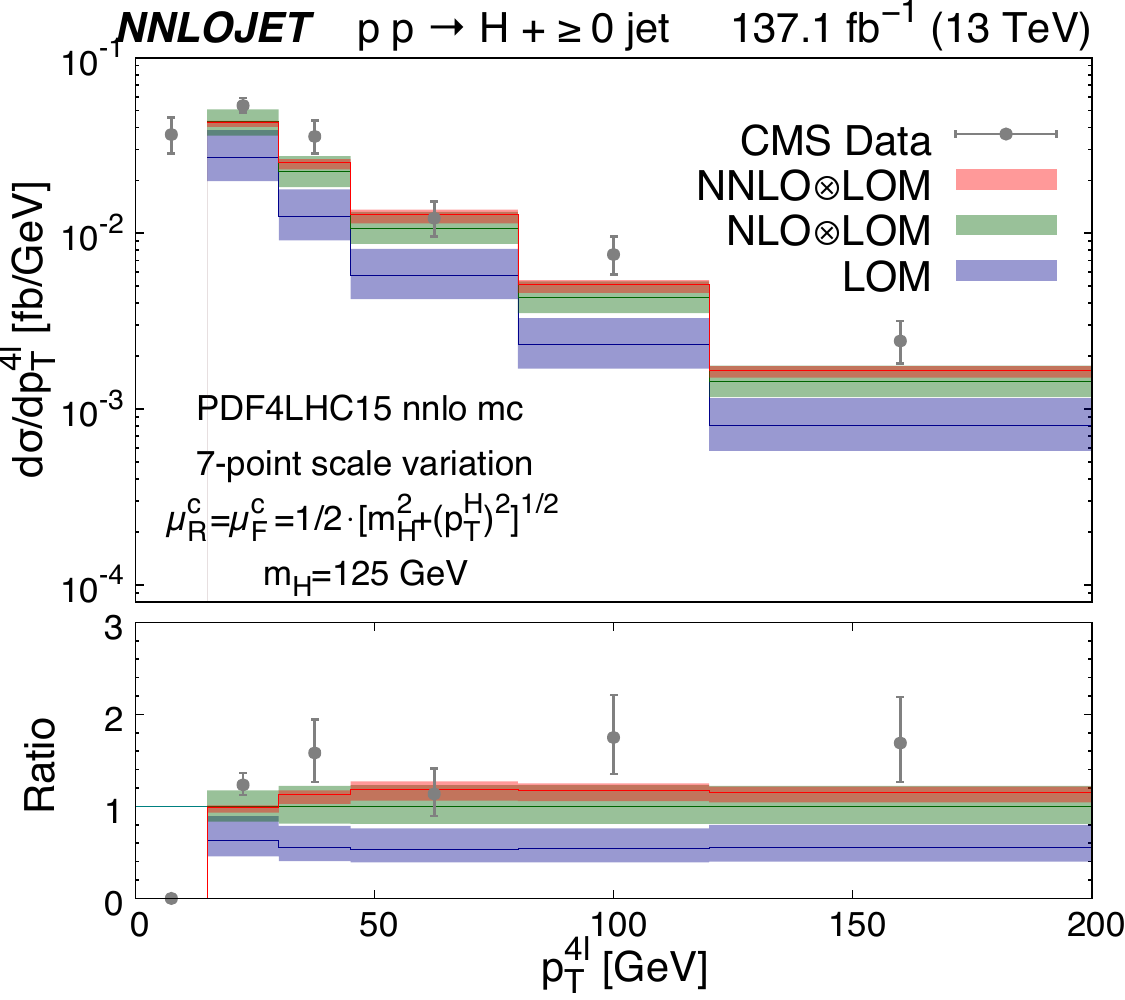}
    \end{subfigure}

    \begin{subfigure}[b]{0.32\textwidth}
        \includegraphics[width=\textwidth]{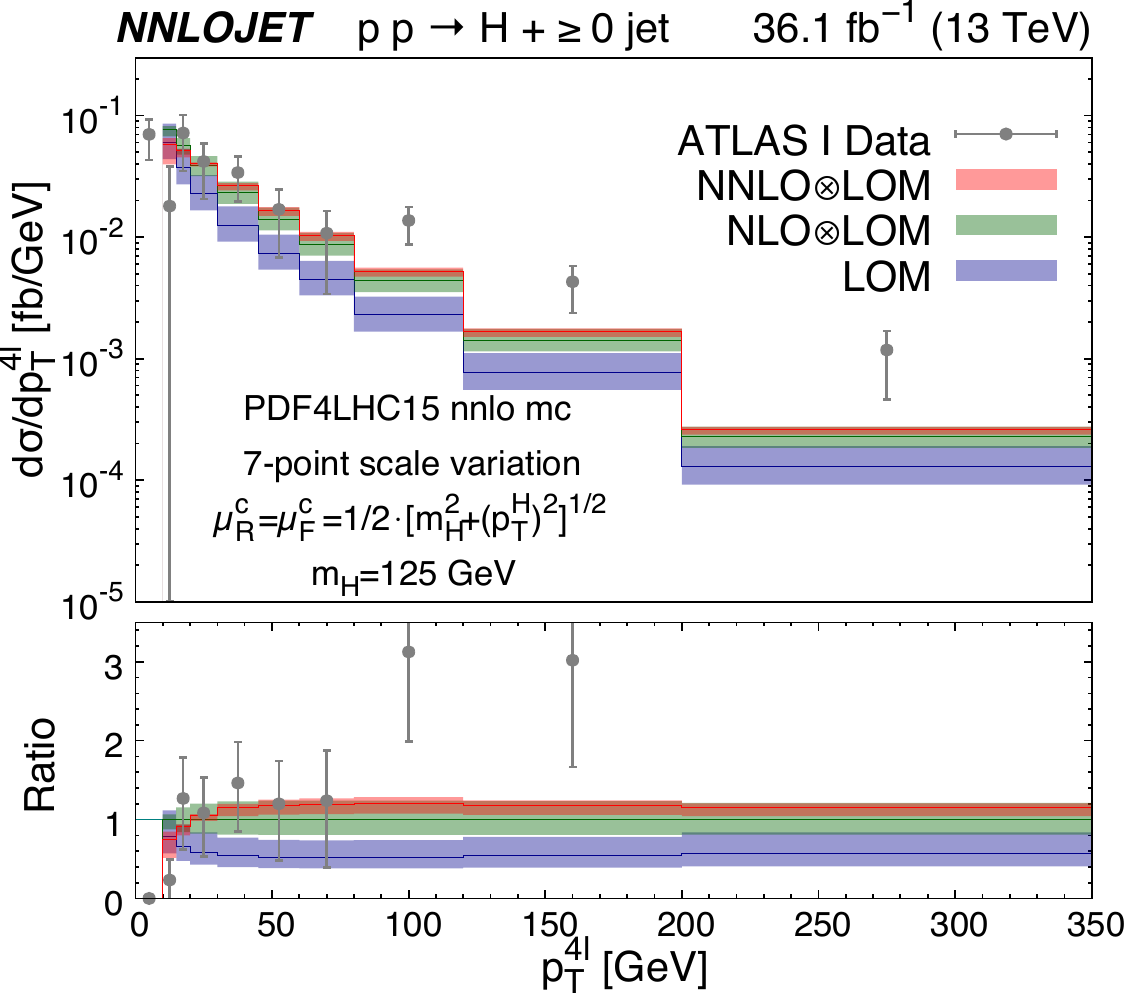}
    \end{subfigure}
    \begin{subfigure}[b]{0.32\textwidth}
        \includegraphics[width=\textwidth]{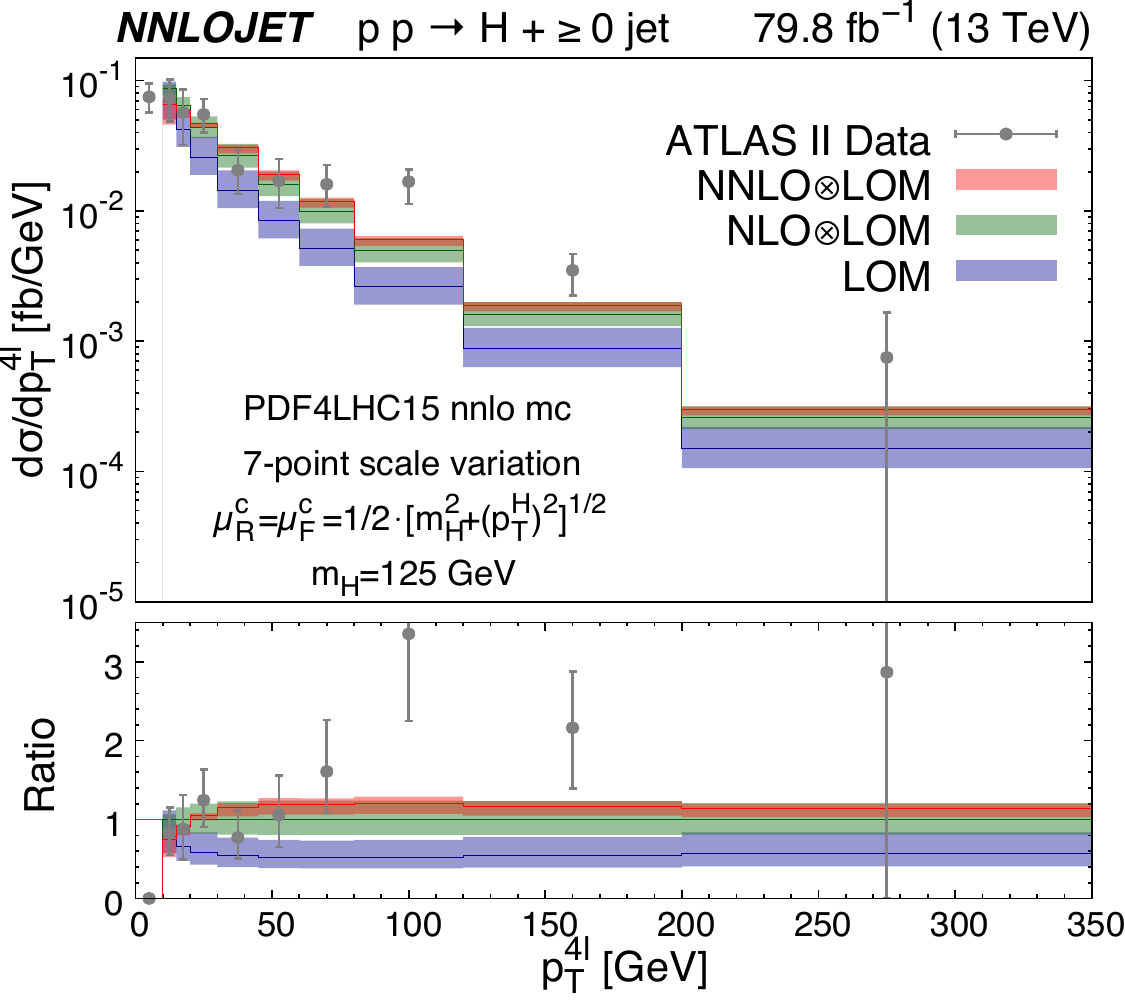}
    \end{subfigure}
    \caption{Transverse momentum distribution of the Higgs boson for CMS [$\int {\cal L}dt =$ 35.9~fb$^{-1}$] (upper-left) and [$\int {\cal L}dt = $ 137.1~fb$^{-1}$] (upper-right), \ATLASI [$\int {\cal L}dt =$ 36.1~fb$^{-1}$] (lower-left) and \ATLASII [$\int {\cal L}dt = $79.8~fb$^{-1}$] (lower-right) cuts and integrated luminosities.}\label{fig:pt4l}
\end{figure}

Higgs boson transverse momentum distributions are compared with the available ATLAS and CMS measurements in Fig.~\ref{fig:pt4l}. The \NNLOJET predictions are obtained from 
the Higgs-plus-jet process by replacing the jet algorithm by a small cut on the Higgs boson transverse momentum, aligned with the first bin edge of the experimental data.  To describe the 
full transverse momentum spectrum including the first bin requires the resummation of large logarithmic corrections~\cite{hresum1,hresum2}.
The NLO and NNLO $K$-factors for these distributions are kinematics dependent and were already discussed above in the 
context of Fig.~\ref{fig:KfactorHpT}. 
We observe that the  NNLO predictions provide an improved description of the experimental data especially 
 for small $p_T^{4l}$ below 80~GeV. At larger transverse momentum, 
the theory description of the  \ATLASII~\cite{ATLASHto4lconf} and CMS~\cite{CMSHto4lconf} data (which each correspond to the larger data sets) is satisfactory within errors. Overall, it is observed 
that the experimental data in this region are typically exceeding the theory predictions, especially for $p_T^{4l}>150$~GeV. In this region, gluon fusion Higgs boson production is complemented by sizeable 
contributions from other production processes (vector boson fusion and 
associated production with a vector boson), which are not included in the present study.

\subsection{Lepton transverse momentum distributions}
\label{sec:leptonpt}

\begin{figure}
    \centering
    \begin{subfigure}[b]{0.4\textwidth}
        \includegraphics[width=\textwidth]{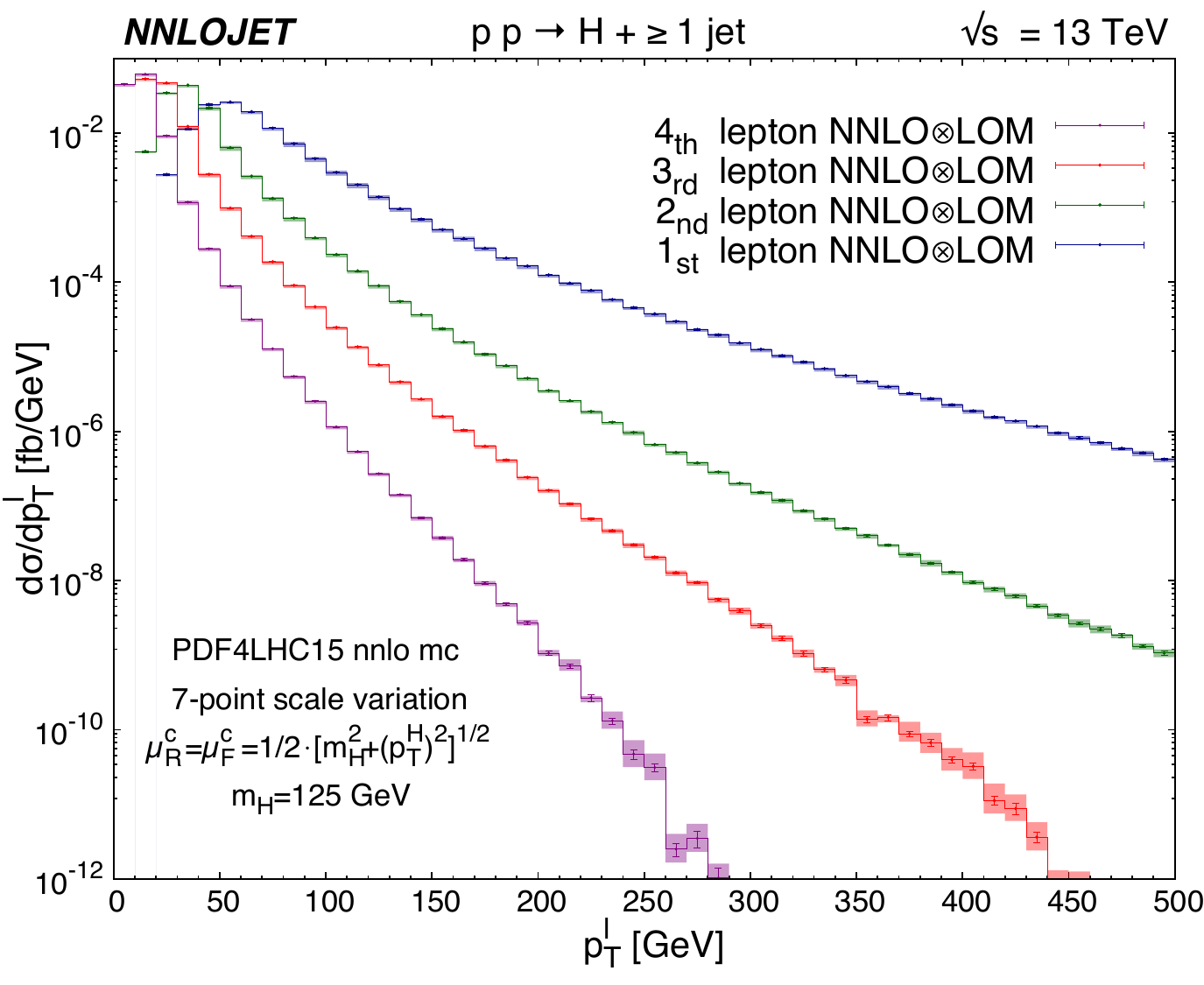}
    \end{subfigure}
    \begin{subfigure}[b]{0.4\textwidth}
        \includegraphics[width=\textwidth]{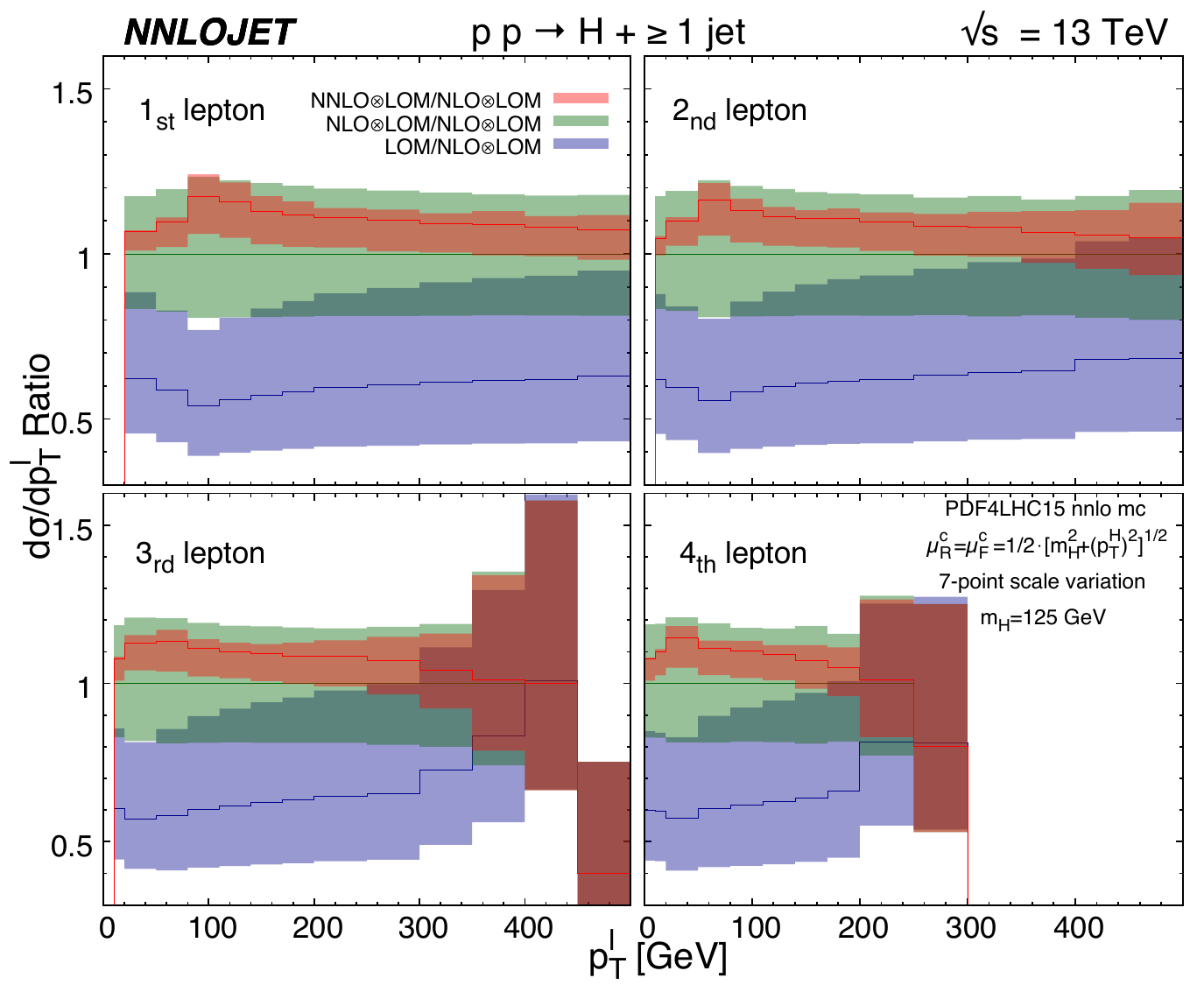}
    \end{subfigure}
    \caption{Transverse momentum distributions for each of the four leptons from Higgs boson decay associated with a jet for \ATLASII cuts.}\label{fig:leptonpt}
\end{figure}

The future increase of the LHC data set will enable multi-differential measurements in specific Higgs-boson final states, thereby opening up novel opportunities for precision studies of Higgs boson production and 
decay. The interpretation of such future measurements will rely on theory predictions for the relevant fiducial distributions. To illustrate that such types of fiducial distributions can be reliably 
predicted using \NNLOJET, we present in Fig.~\ref{fig:leptonpt} the individual transverse momentum distributions of all four leptons in Higgs-plus-jet production, computed for the 
\ATLASII fiducial cuts. The leptons are ordered by their transverse momentum, and summed over charge and flavour. 
The theory uncertainties are obtained from the common seven-point scale variation 
and the numerical integration errors are indicated on the central values as error bars. 
The distributions for the leading and sub-leading leptons are very stable throughout the plotted region. The statistics for the third and fourth leptons drop drastically beyond 300 and 200~GeV respectively, exhibiting relatively large residual integration errors. All four leptons must satisfy the fiducial selection criterion and are identified as isolated leptons. Each of the lepton transverse momentum distribution can be integrated to the same total cross section (within corresponding Monte Carlo errors) which explains the differences in hardness and low-momentum shapes in 
  the left panel of Fig.~\ref{fig:leptonpt}. On the right panel of Fig.~\ref{fig:leptonpt}, predictions for each lepton are normalised to the corresponding NLO$\otimes$LOM distribution. We observe non-flat 
  NNLO corrections for all four leptons with up to  $+17\%$ corrections in the peak region.
   The scale uncertainties for NLO$\otimes$LOM are typically at a level of $\pm 19\%$, which are reduced to about $^{+4\%}_{-8\%}$ at NNLO$\otimes$LOM. These distributions illustrate the non-trivial effects 
   of higher order QCD corrections on fiducial cross sections, which are often controlled by a subtle interplay between fiducial cuts, isolation prescriptions and extra QCD radiation.

\section{Conclusions and outlook}
\label{sec:conc}

In this paper, we investigated the effect of QCD corrections up to NNLO on the fiducial cross sections for Higgs boson production and Higgs-plus-jet production in the $H\to 4l$ decay mode. Owing to its 
kinematic complexity, this decay mode had previously not been computed to NNLO accuracy for Higgs-plus-jet final states and the Higgs boson transverse momentum distribution. Our calculations were 
performed using the \NNLOJET framework in an effective field theory with infinite top quark mass. Finite-mass effects were corrected for by a leading-order multiplicative re-weighting. NNLO corrections 
are found to be sizeable and kinematics dependent, and they lead to a substantial reduction of scale uncertainties to a level of about 10\% in most distributions. 

The four-lepton fiducial cross sections are used frequently to infer simplified template cross sections that enter into global analyses of Higgs boson couplings and properties. This conversion is based on 
acceptance factors, which allow to convert the fiducial measurements to full acceptance. We investigated the perturbative stability of these acceptance factors, and found that they were very stable 
for the ATLAS $H\to 4l$ measurements, while decreasing with increasing order for the CMS  $H\to 4l$ measurements. This difference could be understood to be due to the different lepton isolation prescriptions:
ATLAS uses a jet-lepton overlap criterion, leading to the potential removal of the jet (but not the lepton), while CMS uses an isolation cone around the lepton, leading to a potential removal of the lepton. 
QCD corrections to the acceptance factors are uniform in transverse momentum, but show a dependence on the Higgs boson rapidity in the region approaching the lepton 
rapidity cuts. 

We performed a detailed comparison with the ATLAS and CMS measurements performed at 13~TeV. Overall, the NNLO theory predictions provide a good description of the experimental data, typically 
with uncertainties well below the experimental errors. At large transverse momentum, the experimental measurements usually exceed the theory predictions, indicating the increasing importance of 
other Higgs boson production processes besides the dominant gluon fusion mechanism considered here. 

Our results allow the precise QCD computation of  fiducial cross sections and acceptance factors, and can be extended to arbitrary infrared-safe distributions related to Higgs-plus-jet final states. They 
prepare the ground for precision studies of Higgs boson couplings and properties with upcoming high-luminosity LHC data.

\acknowledgments

We thank Juan Cruz-Martinez, James Currie, Aude Gehrmann-De Ridder,
  Marius H\"ofer, Imre Majer, Jan Niehues, Joao Pires, Duncan Walker and James Whitehead for useful discussions and their many contributions to the \NNLOJET code. 
The authors also thank the University of Zurich S3IT and CSCS Lugano for providing the computational resources for this project. 
This research was supported in part by the UK Science and Technology Facilities Council under contract ST/G000905/1, 
by the Swiss National Science Foundation (SNF) under contract 200020-175595, 
by the Swiss National Supercomputing Centre (CSCS) under project ID UZH10,  
by the Research Executive Agency (REA) of the European Union through the ERC Consolidator Grant HICCUP (614577) and the ERC Advanced Grant MC@NNLO (340983).

\end{document}